\documentclass[sigconf]{acmart}
\renewcommand\footnotetextcopyrightpermission[1]{}
\pdfoutput=1
\DeclareMathDelimiter{(}{\mathopen} {operators}{"28}{largesymbols}{"00}
\DeclareMathDelimiter{)}{\mathclose}{operators}{"29}{largesymbols}{"01}

\usepackage{booktabs}
\usepackage{multirow}
\usepackage{subcaption}
\usepackage{graphicx}
\usepackage{graphbox}
\usepackage{setspace}
\usepackage{wrapfig}
\usepackage{xspace}

\usepackage[colorinlistoftodos,prependcaption,textsize=tiny]{todonotes}
\usepackage{verbatim}

\usepackage{graphicx}
\usepackage{color}
\usepackage{url}
\usepackage{listings}
\usepackage{csquotes}
\usepackage[normalem]{ulem}
\usepackage{setspace}
\usepackage{algorithm}
\usepackage[most]{tcolorbox}
\usepackage[noend]{algpseudocode}

\algblock{Input}{EndInput}
\algblock{Output}{EndOutput}

\usepackage{xparse}
\usepackage{mathtools}
\usepackage{multirow}
\usepackage{adjustbox}
\usepackage{threeparttable} % for table note
\usepackage{enumitem}
\def\BibTeX{{\rm B\kern-.05em{\sc i\kern-.025em b}\kern-.08emT\kern-.1667em\lower.7ex\hbox{E}\kern-.125emX}}

\usepackage{textcomp}
\usepackage{hyperref}

\newcommand{\mybox}[1]{\begin{tcolorbox}[enhanced, frame hidden, boxsep=0pt]\emph{#1}\end{tcolorbox}}

\newcommand{\codeIn}[1]{\texttt{#1}}
\newcommand{\remove}[1]{}
\newcommand{\git}[1]{\textit{GitHub}}

\usepackage{eqparbox}
\newdimen{\algindent}
\algnewcommand\LeftComment[2]{%
\hspace{#1\algindent}$\triangleright$ #2 \hfill %
}

\usepackage{listings}
\usepackage{xcolor}

\definecolor{codegreen}{rgb}{0,0.6,0}
\definecolor{codegray}{rgb}{0.5,0.5,0.5}
\definecolor{uncoveredgray}{rgb}{0.8,0.8,0.8}
\definecolor{codepurple}{rgb}{0.58,0,0.82}
\definecolor{backcolour}{rgb}{0.99,0.80,0.87}

\newcommand{\parabf}[1]{\noindent\textbf{#1}}

\newcommand{\basic}{\textit{K$_0$(Basic)}}

\newcommand{\issuetitle}{\textit{BR(IT)}}
\newcommand{\issuedes}{\textit{BR(ID)}}
\newcommand{\bugreport}{\textit{BR(ALL)}}

\newcommand{\triggertest}{\textit{PI(TT)}}
\newcommand{\errormsg}{\textit{PI(EM)}}
\newcommand{\buggycomment}{\textit{PI(BC)}}
\newcommand{\tec}{\textit{PI(ALL)}}

\newcommand{\kobs}{\textit{K$_1$(CE)}}
\newcommand{\ko}{\textit{K$_1$(PE)}}
\newcommand{\kt}{\textit{K$_2$(PE, PE)}}
\newcommand{\ktbs}{\textit{K$_2$(CE, PE)}}
\newcommand{\kbasic}{\textit{K$_0$(Basic)}}

\newcommand{\ce}{\textit{CE}}
\newcommand{\pe}{\textit{PE}}

\newcommand{\totalcorrect}{300}

\newcommand{\newcorrect}{128}
\newcommand{\multifunctcorrect}{32}

\newcommand{\chatrepair}{\textsc{ChatRepair}\xspace}

\newcommand{\codellama}{\textsc{CodeLlama}\xspace}
\newcommand{\magicoder}{\textsc{Magicoder}\xspace}
\newcommand{\codexedit}{Codex-edit}

\newcommand{\gptturbo}{GPT-3.5-Turbo}
\newcommand{\gptwot}{GPT-3.5}

\newcommand{\defectsj}{Defects4J}
\newcommand{\quix}{QuixBugs}
\newcommand{\dualllm}{Dual-LLM}
\newcommand{\srepair}{\textsc{SRepair}\xspace}

\newcommand{\testfaulure}{test-failure}

\newcommand{\srepairsample}{500}
\newcommand{\srepaircmp}{\textsc{SRepair$_{\srepairsample{}}$}\xspace}
\newcommand{\srepairbsc}{\textsc{SRepair$_{200}$}\xspace}

\newcommand{\repairinfo}{auxiliary repair-relevant information}
\newcommand{\projinfo}{project-specific information}

\newcommand{\plsfix}{plausible fixes}

\newcommand{\aprperf}{repair performance}

\newcommand{\suggestionpatch}{5}

\newcommand{\suggestionmodel}{repair suggestion model}
\newcommand{\patchmodel}{patch generation model}

\usepackage{caption}

\begin{document}
%\DeclareCaptionType{copyrightbox}
% \captionsetup[figure]{font=bf,skip=1pt}%set figure caption
% \captionsetup[table]{font=bf,skip=1pt}%set table caption

\title{How Far Can We Go with Practical Function-Level Program Repair?}

\author{Jiahong Xiang}
\affiliation{
  \institution{Southern University of Science and Technology}
  \city{Shenzhen}
  \country{China}}
\email{xiangjh2022@mail.sustech.edu.cn}

\author{Xiaoyang Xu, Fanchu Kong}
\affiliation{
  \institution{Southern University of Science and Technology}
  \city{Shenzhen}
  \country{China}}
\email{12112620@mail.sustech.edu.cn}
\email{12112822@mail.sustech.edu.cn}

\author{Mingyuan Wu}
\affiliation{
  \institution{Southern University of Science and Technology}
  \city{Shenzhen}
  \country{China}}
\email{11849319@mail.sustech.edu.cn}

\author{Zizheng Zhan}
\affiliation{
  \institution{Kwai Inc.}
  \city{Beijing}
  \country{China}}
\email{zhanzizheng@kuaishou.com}

\author{Haotian Zhang}
\affiliation{
  \institution{Kwai Inc.}
  \city{Beijing}
  \country{China}}
\email{zhanghaotian@kuaishou.com}

\author{Yuqun Zhang}
\affiliation{%
  \institution{Southern University of Science and Technology}
  \city{Shenzhen}
  \country{China}}
\email{zhangyq@sustech.edu.cn}

\begin{abstract}
Recently, multiple Automated Program Repair (APR) techniques based on Large Language Models (LLMs) have been proposed to enhance the repair performance. 
While these techniques mainly focus on the single-line or hunk-level repair, they face significant challenges in real-world application due to the limited repair task scope and costly statement-level fault localization. However, the more practical function-level APR, which broadens the scope of APR task to fix entire buggy functions and requires only cost-efficient function-level fault localization, remains underexplored. %Furthermore, the effectiveness of few-shot learning adopted by previous work to enable function-level APR remains inadequately validated, and the potential of incorporating the \repairinfo{}, such as trigger tests, to further enhance the function-level \aprperf{}, has not been sufficiently investigated. 

% Recently, with the rapid development of Large Language Models (LLMs), multiple LLM-based Automated Program Repair (APR) techniques have been proposed to enhance the repair performance. 

% Therefore, there is an urgent need to extensively study the LLM-based function-level APR and analyze the effect of the \repairinfo{} for further enhancing the \aprperf{}.

In this paper, we conduct the first comprehensive study of LLM-based function-level APR including investigating the effect of the few-shot learning mechanism and the \repairinfo{}. Specifically, we adopt six widely-studied LLMs and construct a benchmark in both the \defectsj{} 1.2 and 2.0 datasets. Our study demonstrates that LLMs with zero-shot learning are already powerful function-level APR techniques, while applying the few-shot learning mechanism leads to disparate repair performance. Moreover, we find that directly applying the \repairinfo{} to LLMs significantly increases function-level \aprperf{}. Inspired by our findings, we propose an LLM-based function-level APR technique, namely \srepair{}, which adopts a dual-LLM framework to leverage the power of the \repairinfo{} for advancing the repair performance. The evaluation results demonstrate that \srepair{} can correctly fix \totalcorrect{} single-function bugs in the \defectsj{} dataset, largely surpassing all previous APR techniques by at least 85\%, without the need for the costly statement-level fault location information. Furthermore, \srepair{} successfully fixes \multifunctcorrect{} multi-function bugs in the \defectsj{} dataset, which is the first time achieved by any APR technique ever to our best knowledge. 

\end{abstract}

\maketitle

%auto-ignore

\section{Introduction}

% \note{importance of APR}
%In the rapidly advancing technological era, software deeply influences every aspect of daily life~\cite{survey-60, survey-61, survey-62}. However, with the expanding scale of software, the incidence of defects is rising, and 
Fixing software defects costs developers a significant amount of time and effort~\cite{survey-1}. To assist developers in reducing the burden of repairing programs, Automated Program Repair (APR) techniques have been proposed to automatically generate potential patches for buggy programs. 
Specifically, the learning-based APR techniques which incorporate the learning power to advance the \aprperf{} have been increasingly studied in recent years. For instance, many such  techniques~\cite{sequencer, dlfix, coconut-xy, recoder_apr, cure, selfapr, rewardrepair} utilize Neural Machine Translation (NMT)~\cite{nmt-fc} such that APR is modeled as a translation task where the objective is to transform buggy code into correct code. %or Large Language Models (LLMs)~\cite{gpt-3-fewshot-learner-fc} from Deep Learning techniques to enhance repair performance. For the NMT-based APR techniques, 
%However, they face challenges such as the difficulty in ensuring the quantity and quality of training data~\cite{alpharepair} and the ineffectiveness of multi-edit repairs~\cite{dear-xy}. To illustrate, the state-of-the-art NMT-based APR tool~\cite{selfapr} only repaired 110 \modify{\sout{in 818 }}bugs of the \defectsj{} dataset~\cite{defects4j}.

More recently, Large Language Models (LLMs) have become largely adopted in various downstream software tasks~\cite{titanfuzz, fuzzgpt, codamosa, libro} including APR~\cite{alpharepair, Repilot, codexapr-xy, fitrepair, chatrepair-xy, CodeLM_on_APR, inferfix} where they have been proven to advance the \aprperf{}~\cite{llm4apr-xy, CodeLM_on_APR, chenapr-xjh, alpharepair, fitrepair}, e.g., the Codex model can fix 32 more bugs than previous APR techniques in the \defectsj{} 1.2 dataset~\cite{llm4apr-xy}. Meanwhile, researchers also propose multiple LLM-based APR techniques~\cite{chatrepair-xy,Repilot,xia2023conversational, alpharepair, inferfix, rap-gen} to further enhance the repair performance. For instance, the state-of-the-art LLM-based APR technique \chatrepair{}~\cite{chatrepair-xy} employs a conversational repair mechanism and successfully fixes 162 out of 337 bugs in a crafted \defectsj{} dataset, causing at least 24.6\% gain compared with all existing techniques. However, many LLM-based APR techniques are proposed for the single-line or hunk-level program repair by auto-completing a single line~\cite{line_apr} or infilling a hunk of code with context~\cite{alpharepair} respectively. They typically rely on identifying statement-level program faults, i.e., given fault locations~\cite{alpharepair, fitrepair, ASE_APR_FT, Repilot, gamma} or applying statement-level fault localization techniques such as Gzoltar~\cite{rap-gen}. Nevertheless, it has been widely argued that accurately identifying statement-level faults can be essentially costly, i.e., demanding fine-grained input or strong assumptions~\cite{fltestcase,finegrainedflinput,flassumption,flsurvey1}, thus potentially limiting the real-world applicability of the single-line or hunk-level APR. On the other hand, the LLM-based function-level APR can be potentially more promising, i.e., applying a generative model such as an LLM to auto-regressively generate the entire patched version of the buggy function by prompting the buggy function into the LLM. To illustrate, first, the function-level APR enables a larger scope of the program repair task---it involves not only the single-line and hunk-level repair tasks, but also a more complicated task which repairs multiple discontinuous lines or hunks within a function~\cite{dissection-xy}. Second, identifying function-level faults tends to be more cost-efficient than identifying statement-level faults, thus rendering the function-level APR more practical in real world~\cite{practitionersexpect,canrefinefl,deepfl}. 
 
% i.e., it is applicable to 2.3× more bugs in \defectsj{} 1.2~\cite{dissection-xy}

While LLM-based function-level APR techniques are more promising, there lacks sufficient study and understanding of them~\cite{llm4apr-xy}, thus potentially hindering the further improved usage of LLMs for APR. Specifically, first, the existing LLM-based APR techniques exhibit significant performance loss for the function-level APR, e.g., incurring a decrease of 33\% in \chatrepair{} and 36.4\% in CodexRepair~\cite{chatrepair-xy, llm4apr-xy} in terms of correctly fixed bugs. Second, the rationale of how such techniques can be affected has not been fully investigated. More specifically, the effectiveness of certain commonly-used mechanism such as the few-shot learning~\cite{llm4apr-xy, inferfix, chatrepair-xy}, i.e., prompting buggy code and fixed code pair examples that illustrate the function-level APR task and provide repair context for advancing the learning power of models, remains inadequately validated. Additionally, the potential of incorporating the \repairinfo{}, such as bug reports and trigger tests, remains underexplored~\cite{chatrepair-xy, codexapr-xy, chenapr-xjh}. 
Thus, there is an urgent need to extensively study the LLM-based function-level APR to further enhance the \aprperf{}.

In this paper, we conduct the first comprehensive study on the function-level LLM-based APR including investigating the effect of the few-shot learning and the \repairinfo{}. Specifically, we adopt six widely-studied LLMs including the state-of-the-art LLMs such as \codexedit{} and \gptturbo{}~\cite{chatrepair-xy, chenapr-xjh, gamma, criticalgptview} as the study subjects. We also construct a benchmark containing 522 single-function bugs, i.e., the bugs existing within one single function, in the \defectsj{} dataset. %for studying the function-level APR across different LLMs. 
%Typically, we build our repair prompt containing the buggy function along with the buggy code and fixed code pair examples provided by the few-shot learning mechanism or the \repairinfo{} for the studied LLMs.
Typically, we build our repair prompt containing the buggy function  along with (1) the buggy code and fixed code pair examples to utilize the few-shot learning mechanism; and (2) the \repairinfo{} such as trigger tests for the studied LLMs respectively. In this way, the entire patched functions can be auto-generated and then validated with the \defectsj{} test suite to derive the plausible patches (which pass all the tests). Our evaluation results demonstrate that incorporating the few-shot learning mechanism in the function-level APR actually causes significantly disparate and even negative impacts on the average number of \plsfix{} %\modify{(i.e., bugs with plausible patches)} 
compared with applying the default LLMs only, i.e., from an increase of 10\% to a decrease of 49.7\% among all studied LLMs. Surprisingly, %we find that only prompting issue title as the \repairinfo{} can yield a substantial 32.2\% improvement on generating \plsfix{} averagely. Meanwhile, 
we find that directly applying trigger tests or error messages for prompting can significantly enhance the repair performance, e.g., 26.7\% and 26.1\% improvement in terms of the average number of \plsfix{} respectively. On the other hand, while statement-level fault location information is shown powerful essential to many APR techniques, only adopting the easier-to-obtain 
\repairinfo{} including trigger tests, error messages, and comments altogether for prompting can achieve rather close performance, i.e., causing merely 7.1\% gap in terms of the number of \plsfix{}. %\jh{On the other hand, while statement-level fault location information is shown powerful and essential to many APR techniques, it shows a non-significant enhancement in repair performance when auxiliary repair-relevant information , such as trigger tests, error messages, and comments, is already provided in combination, i.e., causing merely 8.3\% gap in the average number of plausible fixes compared to additionally adopt statement-level fault location in function-level APR.} 
Such a result indicates the potential of replacing the costly statement-level fault location information for the LLM-based function-level APR. In our study, over 10 million patches are generated and validated, consuming more than 8,000 GPU and 100,000 CPU hours. To our best knowledge, this is the largest empirical study of LLM-based APR conducted to date.

Inspired by our findings, we propose an LLM-based function-level APR technique, namely \srepair{}, which adopts a dual-LLM framework to leverage the power of the \repairinfo{} for advancing the \aprperf{}. In particular, \srepair{} first adopts a \suggestionmodel{} which employs the Chain of Thought (CoT) technique~\cite{fewshotcot} to generate natural-language repair suggestions. More specifically, \srepair{} prompts the LLM with the buggy function and the \repairinfo{} (i.e., trigger tests, error messages, and comments) to identify the root causes of the bugs and generate repair suggestions in natural language accordingly. \srepair{} then adopts a \patchmodel{} to auto-generate a patched function with the assistance of the repair suggestions. Our evaluation demonstrates that \srepair{} can correctly fix a total of \totalcorrect{} single-function bugs in our \defectsj{} dataset, largely surpassing all previous APR techniques, e.g., 1.59$\times$ more than Repilot~\cite{Repilot} and 85\% more than \chatrepair{}~\cite{chatrepair-xy},  without the need for the costly statement-level fault location information. Moreover, \newcorrect{} bugs out of them were not fixed by any of the baseline LLM-based APR techniques adopted in this paper. Surprisingly, \srepair{} is also capable of repairing \multifunctcorrect{} multi-function bugs, i.e., bugs existing across multiple functions at the same time, which, to our best knowledge, is the first time achieved by any APR technique ever. 

%\yyy{highlight the advantage} Specifically, to better leverage the repair capability of \repairinfo{}, \srepair{} adopts a \dualllm{} system that enables: (1) the analysis model to utilize its learning power via Chain of Thought (CoT) technique, analyzing the trigger test, error message and comments with buggy function to generate repair suggestions; (2) the code model to efficiently generate a patched function based on the repair suggestions provided by the analysis model. \yyy{looks like a quite simple approach. is it possible to make it more formal?} In our evaluation, \srepair{} demonstrates superior performance compared to all three state-of-the-art LLM-based APR techniques , achieving \note{} to \note{} more correct fixes in \defectsj{} dataset, and successfully fixed \djonecorrect{} and \djtwocorrect{} bugs in \defectsj{} 1.2 and 2.0 respectively. Remarkably, \srepair{} is also capable of repairing \multifunctcorrect{} multi-function bugs, becoming the new state-of-the-art APR technique. 
% Specifically, \srepair{} is a notably lightweight APR tool, drawing inspiration from defect repair paradigm in the open-source community~\cite{llvm_community}. It employs two LLMs to form a repair system and utilizes Chain of Thought (CoT)~\cite{fewshotcot} techniques to further enhance its capability in analyzing \repairinfo{}.

To summarize, this paper makes the following contributions:
\begin{itemize}
    \item We perform the first ever extensive study on the LLM-based function-level APR with the impact factors on its performance, paving the way for new directions in future research.
    \item We find that LLMs with zero-shot learning are already powerful function-level APR techniques. We also find that applying \repairinfo{} can substantially improve the \aprperf{} for all studied LLMs.
    \item We propose a new LLM-based function-level APR technique, \srepair{}, which can achieve remarkable \aprperf{} by correctly fixing \totalcorrect{} single-function bugs, largely surpassing the SOTA techniques, i.e., outperforming \chatrepair{}~\cite{chatrepair-xy} by 85\% and Repilot~\cite{Repilot} by 1.59$\times$ in the \defectsj{} dataset. Moreover, \srepair{} successfully fixes \multifunctcorrect{} multi-function bugs, which is the first time achieved by any APR technique ever to our best knowledge.
\end{itemize}

%auto-ignore

\section{Background \& Related Work}

\subsection{Large Language Model}

\begin{figure}[htb]
    \centering
    \includegraphics[width = \columnwidth]{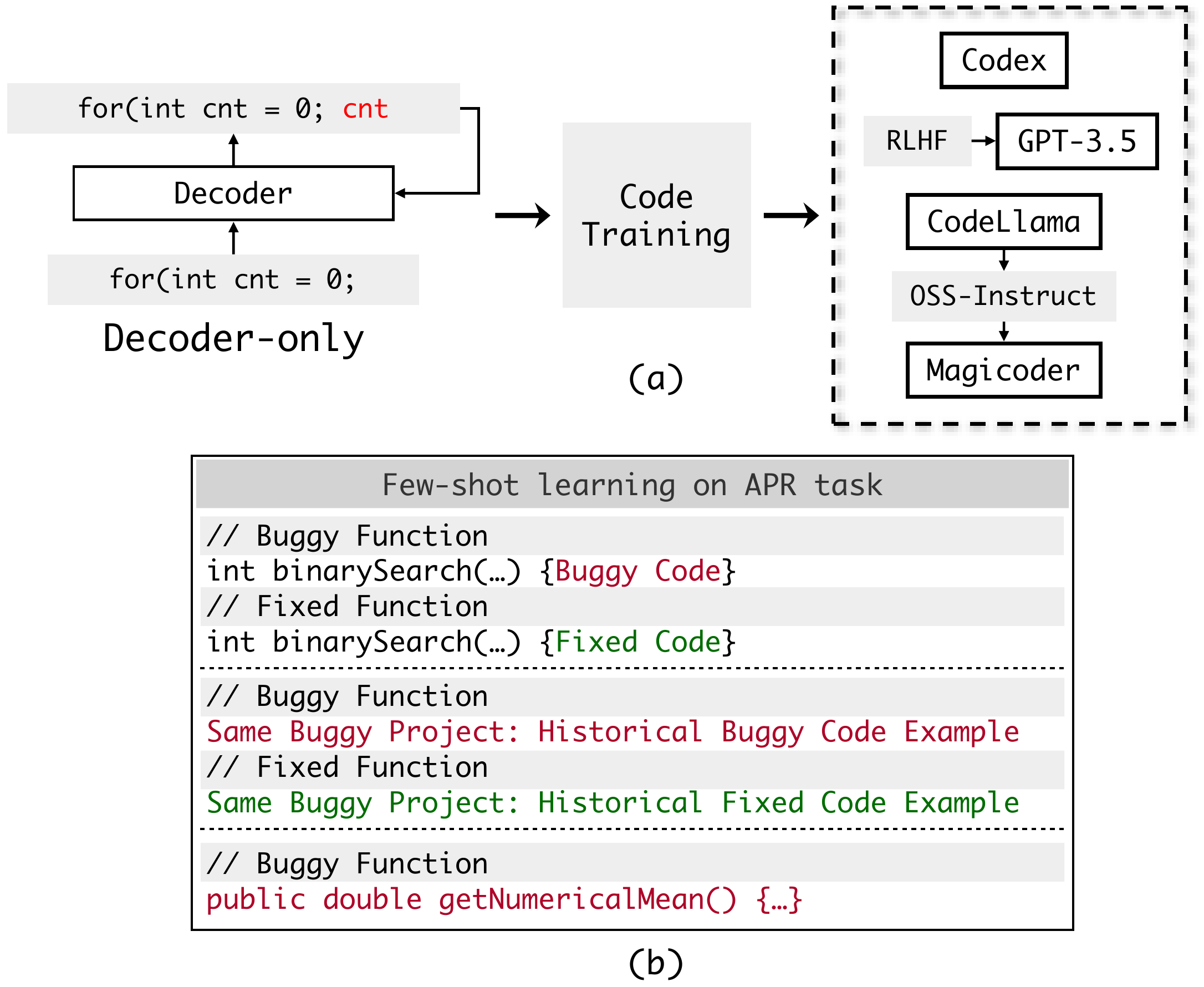}
    \caption{Decoder-only models and few-shot learning on APR}
    \label{fig:llm}
\end{figure}

Large Language Models (LLMs) contain billions of parameters and are trained on petabyte-scale datasets. They are typically built based on the Transformer architecture~\cite{transformer-fc} comprising an encoder for input processing and a decoder for output token generation. %encompassing vast amounts of textual and code data. 
In particular, the decoder-only models as shown in Figure~\ref{fig:llm}a have  demonstrated superior text comprehension~\cite{bert-language-comprehend-fc} and code generation capabilities~\cite{code-gen-fc}. Thus, they have garnered significant interest of researchers and been widely applied to various downstream tasks in software engineering, e.g., test case generation~\cite{libro, codamosa}, vulnerability detection~\cite{titanfuzz, fuzzgpt}, and program repair~\cite{alpharepair, ASE_APR_FT, chenapr-xjh}.  %areas~\cite{alphacode, minecraft, commentgeneration, llmsurvey}. 
These models when integrating domain-specific knowledge for specific tasks, are often fine-tuned~\cite{finetune_bg} %or prompted~\cite{llm_prompt_bg} 
for further improving their performance. For instance, \codellama{}~\cite{codellama-fc} is fine-tuned based on Llama 2~\cite{llama-fc} for generating and discussing code, and \magicoder{}~\cite{mc-on-gh-fc} is fine-tuned with OSS-Instruct~\cite{magicoder-fc} to enhance the code generation performance.
While fine-tuning requires significant computational resources and specialized datasets~\cite{finetune_expensive}, simpler prompting strategies like few-shot learning~\cite{gpt-3-fewshot-learner-fc} and Chain of Thought (CoT)~\cite{fewshotcot} which have also been shown effective are much less costly and thus have been increasingly adopted~\cite{fewshotclinical, cot_example, multimodal_cot_survey}. Figure~\ref{fig:llm}b illustrates how the few-shot learning mechanism is applied for APR. Firstly, the APR task-related examples such as the buggy and fixed code pair of the function \codeIn{binarySearch} are incorporated into the prompt. Note that the example selection varies among different techniques, e.g., manually crafting examples like \codeIn{binarySearch}~\cite{llm4apr-xy} and choosing examples of historical bug fixes within the same buggy project~\cite{chatrepair-xy, llm4apr-xy}. Next, the target buggy function to be fixed, e.g., \codeIn{getNumbericalMean()} in the Math-2 bug~\cite{math-2}, is also added to the prompt. At last, the resulting prompt is fed to the model to generate patches for the target buggy function. To summarize, the purpose of the few-shot learning mechanism is to enable the model to learn how to handle specific tasks through the examples. However, while multiple LLM-based APR techniques have already incorporated the few-shot learning mechanism~\cite{llm4apr-xy, chatrepair-xy, inferfix}, its impacts and characteristics remain unexplored. %Therefore, in RQ1, we aim to investigate the effectiveness and characteristics of few-shot learning mechanism in function-level APR.

\subsection{Automated Program Repair}
Automated Program Repair (APR) techniques~\cite{coconut-xy,alpharepair, cure, Repilot, chenapr-xjh, selfapr, llm4apr-xy, inferfix, ifixr, fitrepair}, designed to aid developers in fixing bugs by automatically generating patches, typically follow the Generate-and-Validate (G\&V) paradigm~\cite{GandV}. In particular, an APR process refers to locating program faults, generating patches for the buggy locations, and validating such patches against a test suite to determine their plausibility (i.e., whether they could pass all tests). Eventually, these resulting plausible patches are manually reviewed to select the correct fix for the target fault. Notably, the trigger tests in the patch-validating test suite are manually created by developers. During the execution of trigger tests, the unit testing framework, e.g., JUnit~\cite{junit}, can be used to provide the corresponding error messages. 

% Moreover, program defects are typically discussed and reported through bug reports, which contain the issue title, i.e., a summary of the bug, and the issue description. Specifically, the issue description includes detailed accounts of the trigger conditions, reproduction steps, and the program's abnormal behavior when the defect is triggered, for the corresponding bugs. However, it is widely argued that the quality of bug reports is inconsistent~\cite{bpQuiality1, bpQuiality2, r2fix}, i.e., they can be invalid, irreproducible, incomplete, or outright misleading. This presents challenges for developers in processing bug reports~\cite{libro} and for some automated techniques~\cite{ifixr} as well.

Among the APR techniques, the learning-based techniques~\cite{selfapr, coconut-xy, cure, dlfix, alpharepair, chatrepair-xy, Repilot} that utilize deep learning techniques have recently achieved remarkable performance. Specifically, many such techniques widely adopt the Neural Machine Translation (NMT)~\cite{nmt-fc} techniques which convert APR into a translation task to transform buggy code into correct code. They typically leverage the power of the NMT models through training on extensive datasets containing millions of buggy and fixed code pairs. However, such techniques are highly costly when building well-constructed datasets of buggy and patch code pairs~\cite{alpharepair} and specific context representation for the NMT models~\cite{cure}. More recently, Large Language Models (LLMs) have become increasingly adopted in various downstream software tasks including APR. In particular, directly applying models like Codex can already outperform all previous APR techniques~\cite{llm4apr-xy}. %achieving 32 more correct fixes than prior state-of-the-art APR tool. 
Meanwhile, multiple LLM-based APR techniques have been proposed to further enhance the \aprperf{}. For instance, AlphaRepair~\cite{alpharepair} applies the pre-trained CodeBERT model with the ``cloze-style'' APR, i.e., removing the buggy code tokens and applying the LLM to generate the correct ones.  Similar as AlphaRepair in adopting the cloze-style repair paradigm, Repilot~\cite{Repilot} focuses on synthesizing compilable patches, utilizing the Language Server Protocol to prune infeasible tokens and proactively complete tokens as suggested by the LLM. FitRepair~\cite{fitrepair} combines LLMs with domain-specific fine-tuning and prompting strategies, fully automating the plastic surgery hypothesis, i.e., the code ingredients to fix bugs usually already exist within the same project. \chatrepair{}~\cite{chatrepair-xy} utilizes the conversational repair mechanism based on \gptwot{} and successfully fixes 162 out of 337 bugs in the \defectsj{} dataset with the assistance of the rich information from original bug-exposing tests.
%Researchers also tend to adopt prompt mechanisms and \repairinfo{} to further enhance the \aprperf{}. 
Fan et al.~\cite{chenapr-xjh} conduct a study to investigate whether APR techniques can correct program errors generated by LLMs, particularly in complex tasks like the LeetCode contests. Another study~\cite{llm4apr-xy} employs the few-shot learning mechanism and  recognizes the ineffectiveness of simply feeding LLMs with only buggy functions as they are not pre-trained for APR. To address this, they create a prompt containing two pairs of buggy and fixed code examples: one manually crafted, and another from the same project or dataset. Then they include the buggy function to be fixed in this prompt, thus activating the function-level APR by providing such a prompt to LLMs. However, their claim that LLMs cannot be directly applied to the function-level APR and the effectiveness of employing the few-shot learning mechanism has not been fully investigated. %Additionally, \chatrepair{} incorporates the rich information from the original bug-exposing tests, such as trigger tests and error messages, into its conversational repair mechanism. Furthermore,  
%Interestingly, they find that the incorporation of fault location information can lead to a decline in repair performance.
% Fakhoury et al.~\cite{nl2fix} construct the Defects4J-Nl2fix dataset, which utilizes natural language repair information from bug report to assess the ability of LLMs in generating correct fixes.

\begin{figure}[htb]
    \centering
    \includegraphics[width = \columnwidth]{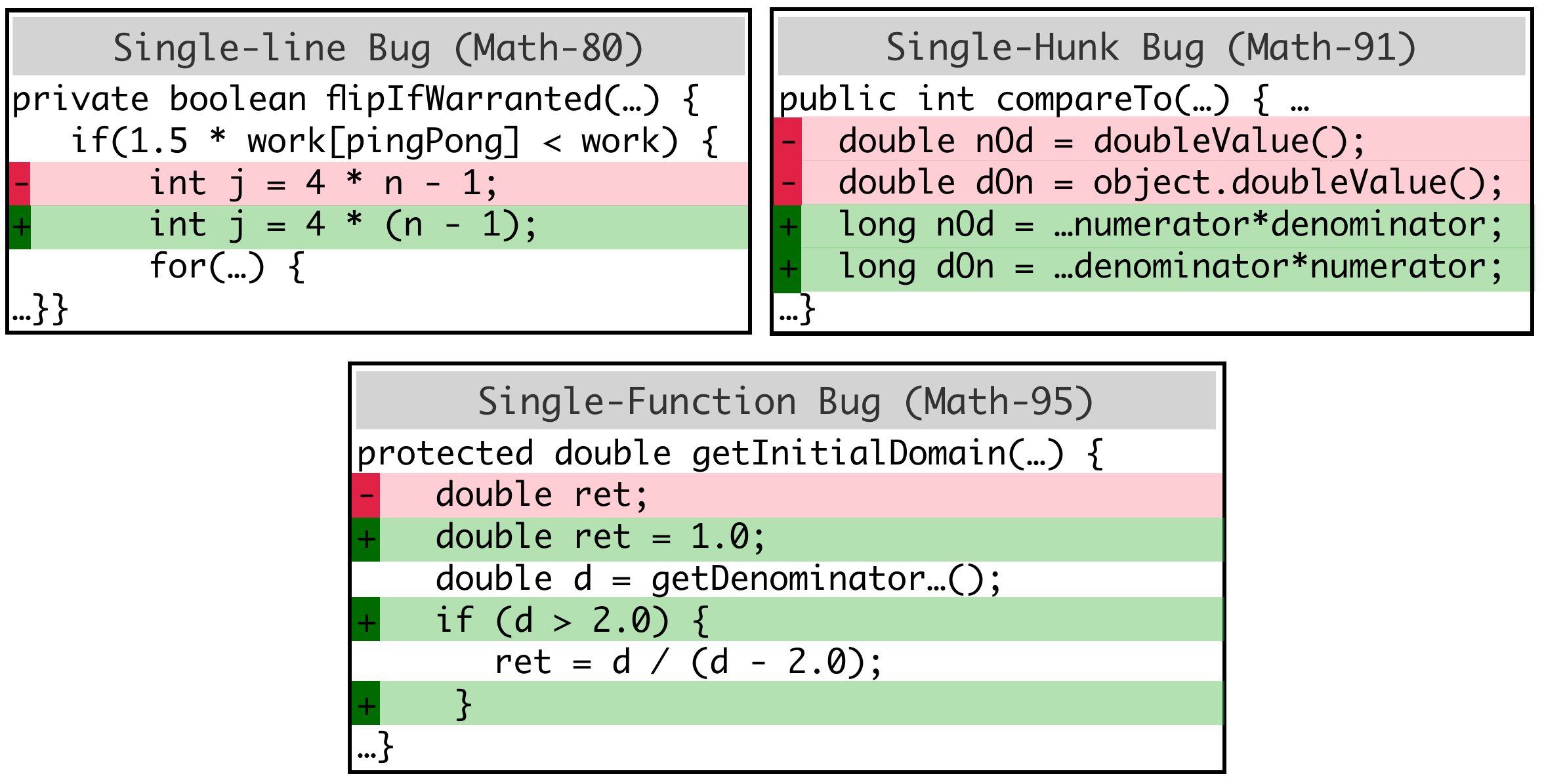}
    \caption{Bug examples existing in a single line, hunk, or function}
    \label{fig:apr-gran}
\end{figure}

% Specifically, Figure~\ref{fig:apr-gran} presents examples of single-line, single-hunk, and single-function bugs. 
The existing LLM-based APR techniques mainly focus on repairing single-line or single-hunk bugs~\cite{alpharepair, fitrepair, Repilot, gamma}, as illustrated in Figure~\ref{fig:apr-gran}. Specifically, the single-line bug Math-80~\cite{math-80} is contained within a single line of the function \codeIn{flipIfWarranted} where fixing this line alone can resolve the bug. Such a single-line bug can be fixed by APR techniques focusing on the line-level repair like AlphaRepair~\cite{alpharepair} given accurate fault locations. Meanwhile, the single-hunk bug Math-91~\cite{math-91} is contained in a continuous section of code where two contiguous buggy lines are replaced in the fixed version. This kind of bugs can be fixed by the hunk-level APR techniques like Repilot~\cite{Repilot} requiring multiple accurate statement-level fault locations. Note that single-line and single-hunk bugs can be considered as part of single-function bugs. On the other hand, the bugs existing in multiple discontinuous sections/lines within a function and requiring simultaneous edits on them for a fix are also referred to as single-function bugs. For instance, fixing the Math-95 bug~\cite{math-95} requires editing three discontinuous lines simultaneously. It can be easily derived that fixing single-function bugs poses a greater challenge for APR techniques, e.g., the state-of-the-art APR techniques like \chatrepair{} and CodexRepair incur a performance decrease of 33\% and 36.4\%~\cite{chatrepair-xy} in terms of the number of correctly fixed bugs for the function-level APR respectively. 

As mentioned, single-function bugs actually include single-hunk and single-line bugs as subset, i.e., enabling a larger scope of repair tasks. % of single-function bugs, and single-line bugs are a subset of single-hunk bugs~\cite{llm4apr-xy, chatrepair-xy}. Consequently, single-function bugs contain all single-line/hunk bugs. 
Moreover, locating function-level bugs tends to be cheaper than locating line-level or hunk-level bugs~\cite{fltestcase,canrefinefl,deepfl}, thus making the function-level APR more practical. %requires statement-level fault localization techniques, which are concerned for the inaccuracy and inconsistency of their results~\cite{fltestcase, finegrainedflinput, flassumption, flsurvey1}. 
Therefore, we consider developing the function-level APR techniques rather promising and worthy being sufficiently investigated. %although function-level APR can be challenging, it covers a broad range of APR tasks, dominating repair tasks over line/hunk-level repair, and only requires more cost-efficient function-level fault localization techniques compared to statement-level ones, thereby making the function-level APR more practical in the real world.

\section{Empirical Study}

\subsection{Study Setup}

\subsubsection{Study Subjects}
We utilize six distinct LLMs as our study subjects, encompassing the widely used state-of-the-art \codexedit{} and \gptturbo{}~\cite{chenapr-xjh, llm4apr-xy, chatrepair-xy, criticalgptview}, along with four advanced open-source code LLMs including \codellama{} 7B, 13B, and 34B (over 500k downloads on Hugging Face within one month~\cite{clm-on-hf-fc}), and \magicoder{} 7B (over 1000 stars on GitHub~\cite{mc-on-gh-fc}). %It is noteworthy that \codellama{} has achieved over 500k downloads on Hugging Face within one month~\cite{clm-on-hf-fc} and is widely regarded as a fundamental open-source code-based LLM~\cite{llmevaluation}, and the \magicoder{} has already garnered over 1000 stars on GitHub~\cite{mc-on-gh-fc} and becomes new state-of-the-art code-based LLM which surpasses the \codellama{} 34B in code related tasks~\cite{magicoder-fc, magicoder_eval}. 
Specifically, we adopt code-davinci-edit-001~\cite{code-davinci-edit-fc}, gpt-3.5-turbo-1106~\cite{gpt-3.5-turbo-1106}, and the CodeLlama-Instruct series~\cite{clm-on-hf-fc} models as the versions of as \codexedit{}, \gptturbo{}, and \codellama{} respectively since they have conducted the instruction fine-tuning~\cite{codellama-fc} and can better follow the APR prompt instruction. We also employ the MagicoderS-CL~\cite{mc-on-gh-fc} model as the version of \magicoder{}. Due to the page limit, the LLM configuration details are presented in our \git{} page~\cite{githubrepo}. 
% Next, we detail the specific models and configurations used in our study.
% In our study, we employ six distinct LLMs for patch generation, which include \codexedit{}, \gptturbo{}, and four open-source LLMs.

% \begin{itemize}
% \item \textbf{code-davinci-edit-001}~\cite{code-davinci-edit-fc} A variant within the Codex family, which supports modifying existing programs and inserting content into programs.

% \item \textbf{gpt-3.5-turbo-1106}~\cite{gpt-3-5-fc} OpenAI's flagship model optimized for chatting, excelling in natural language understanding and code generation.

% \item \textbf{CodeLlama-Instruct}~\cite{codellama-fc} A variant of \codellama{} with enhanced ability in code generation and instruction-following. We use CodeLlama-Instruct with 7B, 13B and 34B parameters.

% \item \textbf{MagicoderS-CL}~\cite{magicoder-fc} A member of \magicoder{} family, which is based on CodeLlama-Python-7B and optimized for code generation.

% \end{itemize}

% Our selection of models, encompassing both high-performing open-source and closed-source code LLMs with parameters ranging from 7B to 34B, ensures comprehensive coverage of use cases and reveals generalizable patterns.

\subsubsection{Dataset}
%auto-ignore
\begin{table}[]
\caption{Statistics of the Dataset}
\label{dataset}
\centering
\setlength\tabcolsep{12pt}
\resizebox{0.95\columnwidth}{!}{%
\begin{tabular}{llr|rr}
\toprule
\textbf{Dataset}                                  & \textbf{Project} & \textbf{\# Bugs} & \textbf{\begin{tabular}[c]{@{}r@{}}SH\\ Bugs\end{tabular}} & \textbf{\begin{tabular}[c]{@{}r@{}}SL\\ Bugs\end{tabular}} \\ \midrule
\multirow{6}{*}{\textit{\textbf{Defects4j 1.2}}}  & Chart            & 16               & 12                                                         & 9                               \\
    & Closure          & 105              & 59                     & 26      \\
  & Lang             & 42               & 23                        & 13           \\
   & Math             & 74               & 35        & 23     \\
    & Mockito          & 24               & 12                   & 7    \\
 & Time             & 16               & 6                 & 3           \\ \midrule
\multirow{11}{*}{\textit{\textbf{Defects4j 2.0}}} & Cli              & 28               & 13                                                         & 6           \\
     & Codec            & 11               & 9                          & 8    \\
       & Collections      & 1                & 1             & 1                \\
       & Compress         & 36               & 16                     & 5         \\
        & Csv              & 12               & 7                      & 4           \\
        & Gson             & 9                & 5                       & 4             \\
     & JacksonCore      & 13               & 9                    & 5             \\
        & JacksonDatabind  & 67               & 26                      & 15                  \\
        & JacksonXml       & 5                & 1               & 1              \\
   & Jsoup            & 53               & 38                          & 27         \\
     & JxPath           & 10               & 4                     & 1            \\ \midrule
\textbf{Overall}          & \textbf{}        & 522              & 276        & 158                                                                                                              \\ \bottomrule
\end{tabular}%
}
\end{table}

We construct our benchmark using both the versions 1.2 and 2.0 of the \defectsj{} dataset \cite{defects4j} which is the most widely used APR dataset~\cite{defects4jexample,cure,llm4apr-xy} with a collection of a total of 835 real-world bugs extracted from open-source Java projects, comprising both buggy and fixed versions of the source code. Notably, we bound our study scope within 522 single-function bugs including 276 single-hunk (``SH'') bugs and 158 single-line (``SL'') bugs as shown in Table~\ref{dataset}. % where ``Bug Tokens'' and ``Change Tokens'' indicate the average number of tokens per bug and the number of tokens altered from buggy to fixed versions respectively. Note that here a token refers to the smallest encoded unit of text~\cite{OpenAIPlatform}. 
It should be noted that our collected single-function bugs already include all the single-line and single-hunk bugs studied in previous works~\cite{llm4apr-xy, alpharepair, fitrepair, chatrepair-xy, Repilot}. %\yyy{how are the single-function and single-hunk concepts correspond to the previous introduction? are they existing concepts or brought up by us only?}
% Each bug in \defectsj{} also contains developer test suite for exposing the defect. 
% Column \textbf{Bug Tokens} and \textbf{Change Tokens} show the average number of tokens of each bug and average number of changed tokens between each buggy and fixed version. 
% Column \textbf{SH Bugs}, \textbf{SL Bugs} show the number of single-hunk bugs and single-line bugs in each project.
% While prior APR techniques using the \defectsj{} dataset often restrict their dataset to contain only single-line, or single-hunk bugs\cite{alpharepair}, function-level repair brings a much larger proportion of bugs into the scope of repair.
% Moreover, the function-level dataset facilitates a more comprehensive exploration of repair scenarios across various bugs, enhancing the potential for thorough research.

% Out of the total 835 bugs documented in both \defectsj{} 1.2 and 2.0, 522 are identified as single-function bugs, forming the benchmark dataset in our study. 
% \note{defects4j 1.2 \& 2.0 per project separately(nl2fix)}
% In all of our experiments we rely on \defectsj{}\cite{defects4j}
% We adopt both two versions of \defectsj{}. \defectsj{} 1.2 contains 391 bugs from 6 open-source Java projects. \defectsj{} 2.0 contains 444 new bugs from 11 additional projects.
% In this paper we specifically targets bugs within single functions. 

% Notably, in this paper, we enable LLM-based function-level APR by providing the LLM with the previously described input prompt in Figure~\ref{fig:init_prompt_setting}. 

\subsubsection{Evaluation Metrics}

To assess the \aprperf{}, we follow the standard practice~\cite{semfix-xy,par-xy,deepfix}, to utilize the plausible patches that pass all test cases as our major evaluation metric. In particular, those test cases include the trigger tests in the \defectsj{} dataset designed to expose bugs and relevant tests which can load the classes associated with the buggy functions.

% that load the class containing the buggy function under test.

% The \defectsj{} dataset includes trigger tests to expose bugs and relevant tests that load the class containing the method under test. 
% To assess repair performance, we employ standard practice\cite{semfix-xy,par-xy,deepfix}, focusing on plausible patches that pass all test cases. 
% In our study, a patch is considered to be plausible only if it passes both the trigger and relevant tests.

\subsection{Research Questions}

We investigate the following research questions for extensively studying the function-level APR along with the factors which can impact its effectiveness. %few-shot learning mechanism and \repairinfo{} in LLM-based function-level APR. 

\begin{itemize}[leftmargin=*]

    \item \parabf{RQ1:} \textit{How does the LLM-based function-level APR perform under the zero-shot and few-shot learning setups?} For this RQ, we attempt to investigate the performance of the LLM-based function-level APR under the default zero-shot learning and explore the performance impact from adopting few-shot learning.
    \item \parabf{RQ2:} \textit{How do different \repairinfo{} affect the LLM-based function-level \aprperf{}}? For this RQ, we attempt to study the impact from different \repairinfo{} including bug reports, trigger tests, etc., on the function-level \aprperf{}.
\end{itemize}

%\modify{\sout{\textbf{RQ1:} How does the few-shot learning mechanism, which applies buggy code and fixed code pair examples to illustrate the APR task and provide repair context, perform in function-level APR across different LLMs? For this RQ, we attempt to investigate the performance impact of the few-shot learning mechanism used in previous works on function-level APR.}}

% \item \textbf{RQ3:} What is the relationship between repair scenarios containing implicit defect location and statement-level fault localization information? For this RQ, we investigate the relationship and the impact of the statement-level fault localization information by comparing with other repair scenarios.
% Considering that different RQs adopt various repair scenarios, we introduce them into each respective RQ for a more clear presentation of our setting.
% Considering that different RQs employ various repair scenarios, we detail these scenarios within their respective RQs to clarify our specific experimental setup. 

% \subsection{Result Analysis}
\subsection{Implementation}
% \note{temperature, sample size}
% \note{server CPU GPU Ubuntu OpenJDK}
% \note{\basic{} prompt setting}

We obtain the model from Hugging Face~\cite{hugging-face-fc} and access \codexedit{} and \gptturbo{} through API~\cite{openai-api-fc} provided by OpenAI. Our default setting for patch generation uses the nucleus sampling with top $p$ = $0.9$, temperature = $0.8$ and $200$ samples per bug following prior works~\cite{codexapr-xy,llm4apr-xy,codex-fc}. Patches are generated on servers with 128-core 2.6GHz AMD EPYC™ ROME 7H12 CPU, 512 GiB RAM and eight NVIDIA A100 80GB GPUs, running Ubuntu 20.04.6 LTS. 

\subsubsection{APR input prompt setup}
\begin{figure}[!htb]
    \centering
    \includegraphics[width=0.93\columnwidth]{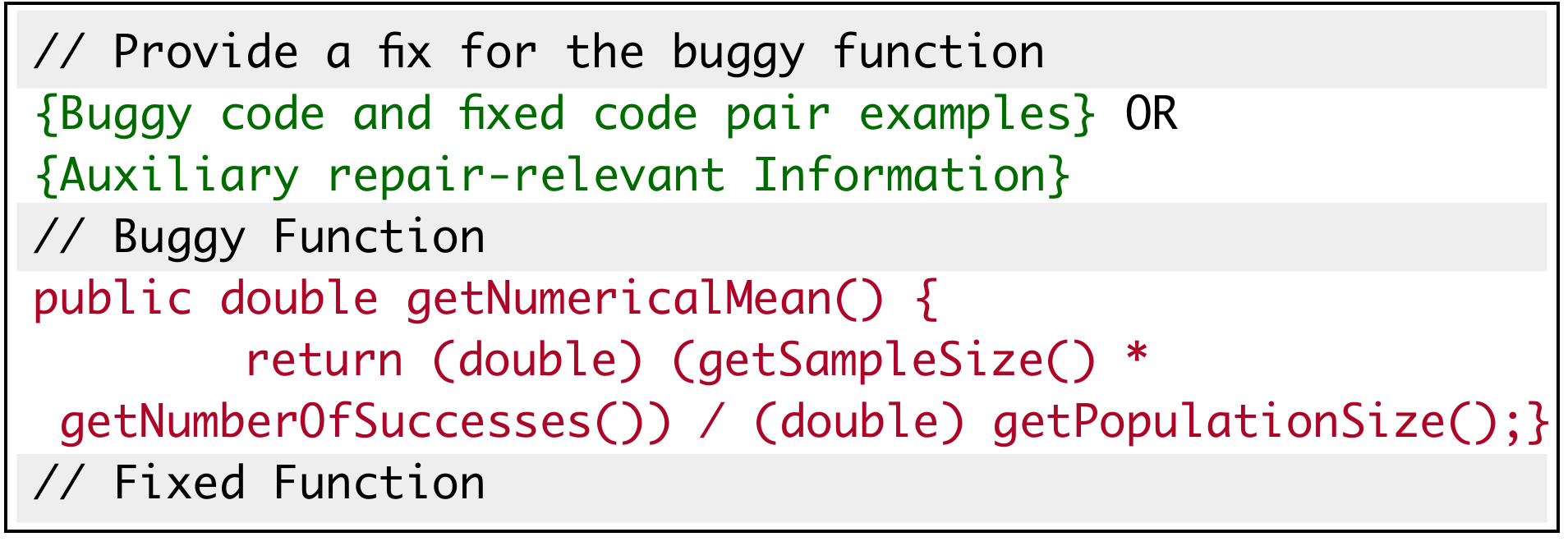}
    \caption{The input prompt for the function-level APR of the Math-2 bug}
    \label{fig:init_prompt_setting}
\end{figure}

Following prior studies~\cite{llm4apr-xy, codexapr-xy}, we set the prompt for the LLM utilized in the APR task, as illustrated in Figure~\ref{fig:init_prompt_setting} to enable the function-level APR. Specifically, we begin with a description of the APR task as `\codeIn{Provide a fix for the buggy function}'. Next, we incorporate the buggy code and fixed code pair examples from the few-shot learning mechanism or the \repairinfo{} into the prompt. Subsequently, we use `\codeIn{Buggy Function}' in conjunction with the buggy code, e.g., the Math-2 buggy function \codeIn{getNumericalMean}~\cite{math-2} in Figure~\ref{fig:init_prompt_setting}, to prompt LLMs with the buggy function to be fixed. Finally, we apply `\codeIn{Fixed Function}' to guide the LLM in generating a patched function. Notably, we employ the zero-shot learning approach as the default baseline in our study, i.e., adopting no \repairinfo{} or buggy-fixed code pair examples for prompting.

\subsubsection{K-shot learning setups}
%auto-ignore
\begin{table}[]
\caption{K-shot learning settings and abbreviations}
\centering
\label{fs-setting}
\resizebox{0.95\columnwidth}{!}{%
\setlength\tabcolsep{12pt}
\begin{tabular}{c|c|c}
\toprule
\textbf{K} & \multicolumn{1}{c|}{\textbf{Example Type}} & \textbf{Abbreviation} \\
\midrule
0 & \multicolumn{1}{c|}{N.A.} & \textbf{\kbasic{}} \\

1 & \multicolumn{1}{c|}{Crafted Example} & \textbf{\kobs{}} \\

1 & \multicolumn{1}{c|}{Project Example} & \textbf{\ko{}} \\

2 & \multicolumn{1}{c|}{Crafted Example \& Project Example} & \textbf{\ktbs{}} \\

2 & \multicolumn{1}{c|}{Project Example \& Project Example} & \textbf{\kt{}} \\
\bottomrule
\end{tabular}%
}
\end{table}

% \begin{table}[]
% \caption{K-shot Settings in RQ1}
% \centering
% \label{fs-setting}
% \resizebox{0.8\columnwidth}{!}{%
% \begin{tabular}{c|c|c}
% \toprule
% \textbf{K} & \multicolumn{1}{c|}{\textbf{Example Type}} & \textbf{Abbreviation} \\
% \midrule
% \textbf{0} & \multicolumn{1}{c|}{\textbf{N.A.}} & \textbf{\basic{}} \\

% \textbf{1} & \multicolumn{1}{c|}{\textbf{Crafted Example}} & \textbf{\kobs{}} \\

% \textbf{1} & \multicolumn{1}{c|}{\textbf{Project Example}} & \textbf{\ko{}} \\

% \textbf{2} & \multicolumn{1}{c|}{\textbf{Crafted Example \& Project Example}} & \textbf{\ktbs{}} \\

% \textbf{2} & \multicolumn{1}{c|}{\textbf{Project example \& Project Example}} & \textbf{\kt{}} \\

% \bottomrule

% \end{tabular}%
% }
% \end{table}
Table~\ref{fs-setting} presents our k-shot learning setups. Specifically, we set the zero-shot learning approach, i.e., adopting no pairs of buggy code and fixed code examples (K=0), as our basic setup denoted as \kbasic{}. Moreover, we follow prior work~\cite{llm4apr-xy} to form our buggy and fixed code pair examples via manually crafted examples (\ce{}), i.e., \codeIn{binarySearch}  in Figure~\ref{fig:llm}b and chosen historical bug fix examples within the same buggy project (\pe{}). We thus form our k-shot learning setup variants as \ko{} with only one chosen historical bug fix example from the same project, \kobs{} with only one manually crafted example, and \ktbs{} with one manually crafted example and one chosen historical bug fix example from the same project. We also form \kt{} with two chosen historical bug fix examples from the same project following the implementation of the prior work for selecting multiple examples~\cite{multi-example-impl}. Note that while it is possible for more setup variants, e.g., with more manually crafted examples and chosen historical bug fix examples from the same project, we generally follow the setup of prior work~\cite{llm4apr-xy} for fair performance comparison and cost-efficient evaluations. 

\subsubsection{Collecting \repairinfo{}}

\begin{figure}[!htb]
    \centering
    \includegraphics[width=0.98\columnwidth]{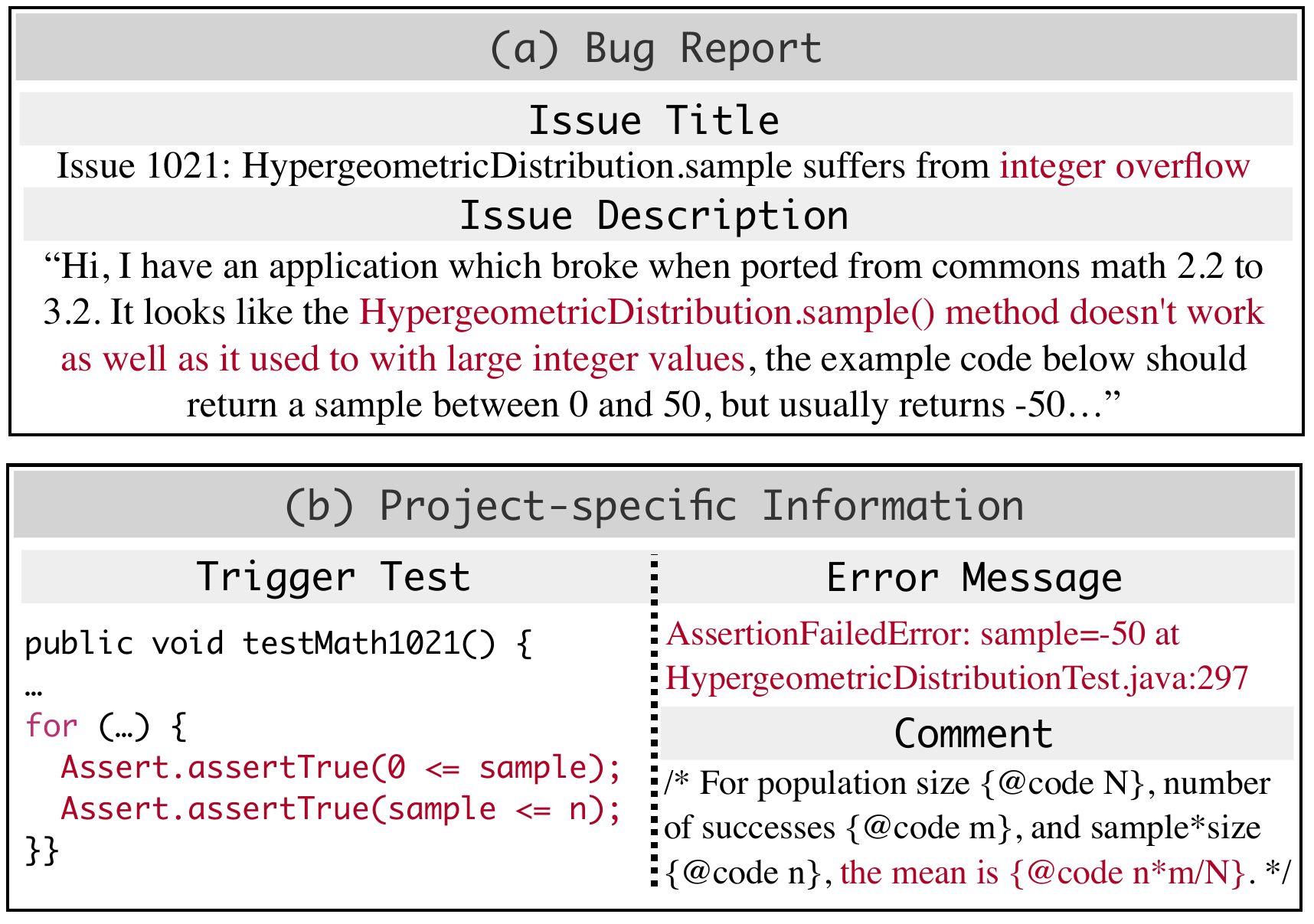}
    \caption{The bug report and \projinfo{} in the Math-2 bug}
    % The \repairinfo{} of bug report and \projinfo{}, exemplified by Math-2 bug
    \label{fig:bp-at}
\end{figure}

In this study, we refer to the \repairinfo{} as the bug report and \projinfo{} from the target buggy project following prior works~\cite{elixir, chatrepair-xy,selfapr,iter, ifixr}, as the Math-2 bug shown in Figure~\ref{fig:bp-at}. Specifically, the bug reports are collected from the official issue links of the \defectsj{}~\cite{d4j_bugreport} repository. More specifically, following prior works~\cite{bugreport_div, bugreport_des}, we divide a bug report into two parts, as illustrated in Figure~\ref{fig:bp-at}a. One is the issue title with averagely around 12 tokens to summarize the type and the cause of the bug (e.g., \textit{``Issue 1021: ... suffers from integer overflow''}). The other is the issue description with averagely 234 tokens which provides detailed conditions, error messages, and reproduction steps, etc. For instance, the issue description in the Math-2 bug report provides a detailed description of the buggy method \codeIn{HypergeometricDistribution.sample()} with the trigger conditions, i.e., \textit{``with large integer values''}.

Furthermore, we automatically extract the \projinfo{} from the buggy project, as the Math-2 bug in Figure~\ref{fig:bp-at}b following the prior works~\cite{chatrepair-xy,selfapr,codexapr-xy}. Specifically, we first build all the buggy projects and automatically extract all the trigger tests and buggy function comments. Then, for each bug, we execute the trigger tests and capture the error messages generated by the unit test framework, such as Junit~\cite{junit}. %These trigger tests contain the trigger conditions of the bugs, e.g., \codeIn{assertTrue(0<=sample)}. Additionally, the error message implicitly includes the boundary conditions that trigger defects, such as \codeIn{sample=-50}. Interestingly, we find that the functionality description is contained in comments, for example, \textit{`the mean is n*m/N'} in the Math-2 bug. 
Notably, among all 522 single-function bugs in the \defectsj{} dataset, only 10 miss reports and 2 miss comments. We then leave such \repairinfo{} empty in our study.

% \paragraph{\textbf{Evaluation Setup}}
For evaluating the impact from the \repairinfo{}, we form eight different setups. Specifically, for bug report-relevant information, we form three setups: \issuetitle{} with the issue title only, \issuedes{} with the issue description only, and \bugreport{} with the whole bug report. For the \projinfo{}, we form four setups: \triggertest{} with the trigger test only, \errormsg{} with the error message only, \buggycomment{} with the buggy comment only, and \tec{} with all such information.   %It should be noted that all these settings are incorporated into the corresponding part of the APR input prompt and then provided to the LLM for generating the patched function.  \yyy{so many variables. reviewers would wonder which is actually important}

\subsection{Result Analysis}

%auto-ignore
\begin{table*}
    \caption {APR results under different few-shot learning setups}
    \label{tab:rq1-few-shot-result}
    \centering
    \setlength\tabcolsep{18pt}
    \begin{adjustbox}{width=0.99\textwidth}
    \begin{tabular}{lrrrrrrr}
    \toprule
\multirow{2}{*}{\textbf{Settings}} & \multirow{2}{*}{\bf{\codexedit{}}} & \multirow{2}{*}{\bf{\gptturbo{}}} & \multicolumn{3}{c}{\bf{\codellama{}}} & \multirow{2}{*}{\bf{\magicoder{}}} & \multirow{2}{*}{\textit{\textbf{Average Plausible Fixes}}} \\

    \cline{4-6}
    & & & \textbf{7B} & \textbf{13B}  & \textbf{34B} \\
    \hline
    %                  codex    gpt       c7b           c13b            c34b    magic           avg
    \bf{\kbasic{}}   & \bf{174} &   175  &   192     &   179      &   160    &   \bf{199}   &   \bf{180} \\ 
    \bf{\kobs{}}    &   103  &   138  &   180       &   185      & \bf{176} &   112        &   149 \\ 
    \bf{\ko{}}      &   109  &   174  &   \bf{194}  &   \bf{193} &   153    &   157        &   163 \\ 
    \bf{\ktbs{}}    &   138  &   166  &   175       &   189      &   125    &   100        &   149 \\ 
    \bf{\kt{}}      &   165  &  \bf{187}  &   167   &  189       &   128    &   121        &   160 \\ 
    % \hline
    % \bf{Average}      &   138  &  168  &   182   &  187       &   148    &   138        &   160 \\ 
    \bottomrule
    \end{tabular}
    \end{adjustbox}
\end{table*}

\subsubsection{RQ1: the function-level \aprperf{}}

Table~\ref{tab:rq1-few-shot-result} presents the function-level APR results in terms of the number of plausible fixes. In general, we observe that \kbasic{} achieves the overall optimal plausible fix results, i.e., 180 average \plsfix{} out of our collected 522 single-function bugs, outperforming all the rest setups by at least 10.4\%. %Notably, \magicoder{} stands out by generating 199 \plsfix{}. 
Such a result indicates that LLMs themselves (with zero-shot learning) are already powerful function-level APR techniques. %More specifically, we can observe that five code models consistently achieve surprising performance in the \basic{} setting, with this performance ranging from 160 \plsfix{} in \codellama{} 34B to 199 in \magicoder{}. Interestingly, we find that while the state-of-the-art conversational model, \gptturbo{}, obtains 175 \plsfix{} in the \basic{} setting, the other five code models achieve an average of 180.4 \plsfix{}. Such result indicates that a code model can achieve impressive performance on function-level APR, even in the small model that only has 7B parameters.

\mybox{Finding 1: LLMs with zero-shot learning are already powerful function-level APR techniques.}
% LLMs themselves are powerful function-level APR techniques in zero-shot learning setting

% To summarize, we find that in zero-shot learning setting, LLMs themselves are powerful function-level APR techniques.
% However, in \codellama{} models, we can find that repair performance is declined with growing of parameter size, i.e., 192 \plsfix{} in 7B model, 179 \plsfix{} in 13B model and only 160 \plsfix{} in 34B model.

Interestingly, we can further observe that applying the few-shot learning mechanism leads to quite disparate plausible fix results across LLMs. For instance, compared with \kbasic{}, while \codellama{} 34B shows a 10\% (176 vs. 160) improvement in \kobs{}, \magicoder{} shows a 49.7\% (100 vs. 199) decline in terms of the number of plausible fixes.

\mybox{Finding 2: Applying the few-shot learning mechanism in the function-level APR leads to disparate plausible fix results across LLMs.}
% \mybox{Finding 2: The few-shot learning mechanism does not effectively enable LLM-based function-level APR and may even negatively impact the repair performance.}

% \mybox{Finding 2: While an LLM on its own is a powerful APR technique, the additional adoption of a few-shot learning mechanism does not effectively enable function-level APR and can even negatively impact the repair performance.}

\begin{figure}[!htb]
    \centering
    \includegraphics[width=\columnwidth]{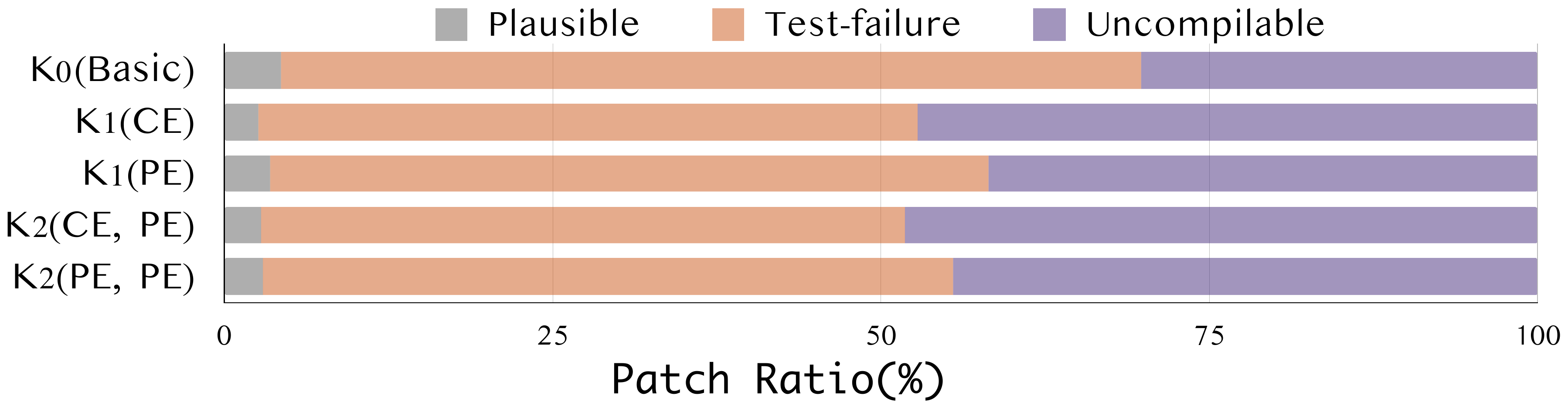}
    \caption{Patch status averaged across all models under different few-shot learning setups}
    \label{fig:fs-pe}
\end{figure}

Furthermore, we present the distribution of plausible, \testfaulure{}, and uncompilable patches given the identical total number of generated patches across different LLMs under all setups in Figure~\ref{fig:fs-pe}. Note that the \testfaulure{} patches can be successfully compiled but fail one or more tests in our \defectsj{} dataset. Interestingly, we can find that \kbasic{} achieves the best plausible patch rate 4.3\% and the lowest uncompilable patch rate 30.2\% among all the k-shot setups, while applying the few-shot learning generates more uncompilable patches, i.e., ranging from 38.4\% to 59.6\% more than \kbasic{}. %Moreover, we observe that few-shot learning settings incorporating the crafted example (\ce{}) result in a more pronounced increase in the uncompilable ratio compared to those incorporating the problem example (\pe{}), with \kobs{} rising to 47.2\% compared to 41.8\% in \ko{}, and \ktbs{} increasing to 48.2\% versus 44.5\% in \kt{}. Such result indicates that including the crafted example exacerbates the uncompilable patch ratio, consequently diminishing repair performance. Interestingly, we also observe that the ratio of uncompilable code slightly increases as the number of examples (K) increases. For instance, \ko{} and \kobs{} have uncompilable ratios of 41.8\% and 47.2\% respectively, while \kt{} and \ktbs{} exhibit higher uncompilable ratios of 44.5\% and 48.2\% respectively. Therefore, we can infer that as the number of examples increases, the uncompilable ratio also tends to rise.

\mybox{Finding 3: Applying the few-shot learning mechanism may generate more uncompilable patches than the zero-shot learning mechanism.}

\subsubsection{RQ2: performance impact from the \repairinfo{}}

% Then we automatically extract the \projinfo{}, i.e., trigger test, error messages and comments, from the buggy project to discuss their performance impact used as \repairinfo{} for function-level APR. Finally, while the function-level APR does not required the statement-level fault location information, we consider such information as the \repairinfo{} to further discuss the performance impact and necessity on function-level APR.
%\subsubsection{The \repairinfo{}}

% \paragraph{\textbf{Result Analysis}}

%auto-ignore
\begin{table*}
    \caption {APR result in different \repairinfo{} settings}
    \label{tab:rq2-bpat-result}
    \setlength\tabcolsep{16pt}
    \begin{adjustbox}{width=0.99\textwidth}
    % \begin{threeparttable}
    \begin{tabular}{p{2cm}lrrrrrrrr}
    \toprule
    % \hline
\multirow{2}{*}{\textbf{Sources}}&\multirow{2}{*}{\textbf{Settings}} & \multirow{2}{*}{\bf{\codexedit{}}} & \multirow{2}{*}{\bf{\gptturbo{}}} & \multicolumn{3}{c}{\bf{\codellama{}}} & \multirow{2}{*}{\bf{\magicoder{}}} & \multirow{2}{*}{\textit{\textbf{Average Plausible Fixes}}} \\

    \cline{5-7}
    & & & &\textbf{7B} & \textbf{13B}  & \textbf{34B} \\
    % \hline
    \midrule      
        %                                                   codex       gpt         c7b     c13b    c34b        magic       avg
       \bf{N.A.} & \bf{\kbasic{}}                        &   174     &   175     &   192  &   179  &   160  &   199 &   180 \\
       \cmidrule(lr){1-9}   
       \multirow{3}{2cm}{\textbf{Bug Report Information}}
        &\bf{\issuetitle{}}                             &   265     &   233     &   234  &   221  &   221  &   251      &   238 \\ 
        &\bf{\issuedes{}}                               &   281     &   \bf{286}&   261  &   \bf{264}  &   248  &   \bf{279}      &   270 \\ 
        &\bf{\bugreport{}}           &   \bf{301}&   285     & \bf{275}  &   260  &   \bf{255}  &   260      &   \bf{273} \\       
        \cmidrule(lr){1-9} 
    \multirow{4}{2cm}{\textbf{Project-specific Information}}
        &\bf{\buggycomment{}}                           &   186  &   185        &   187  &   191  &   169  &   194 &   185 \\ 
        &\bf{\errormsg{}}                               &   217  &   226        &   239  &   225  &   217  &   240 &   227 \\ 
        &\bf{\triggertest{}}                            &   239  &   247        &   227  &   221  &   201  &   235      &   228 \\ 
        &\bf{\tec{}}    &   \bf{264}& \bf{273}  &\bf{249} &  \bf{247}&   \bf{236}  & \bf{254}      &   \bf{254} \\ 
        % \hline
        \bottomrule
        \end{tabular}
        % \begin{tablenotes}
        % \item[{}] \textbf{\bugreport{}\textsuperscript{†} combines \issuetitle{} and \issuedes{}.}
        % \item[{}] \textbf{\tec{}\textsuperscript{\textdaggerdbl} represents the collective usage of \triggertest{}, \errormsg{}, and \buggycomment{}.}
        % \end{tablenotes}
        % \end{threeparttable}
        \end{adjustbox}
\end{table*}

 \begin{figure}
    \centering
    \begin{subfigure}[t]{1.25in}
        \centering
        \includegraphics[width=\textwidth]{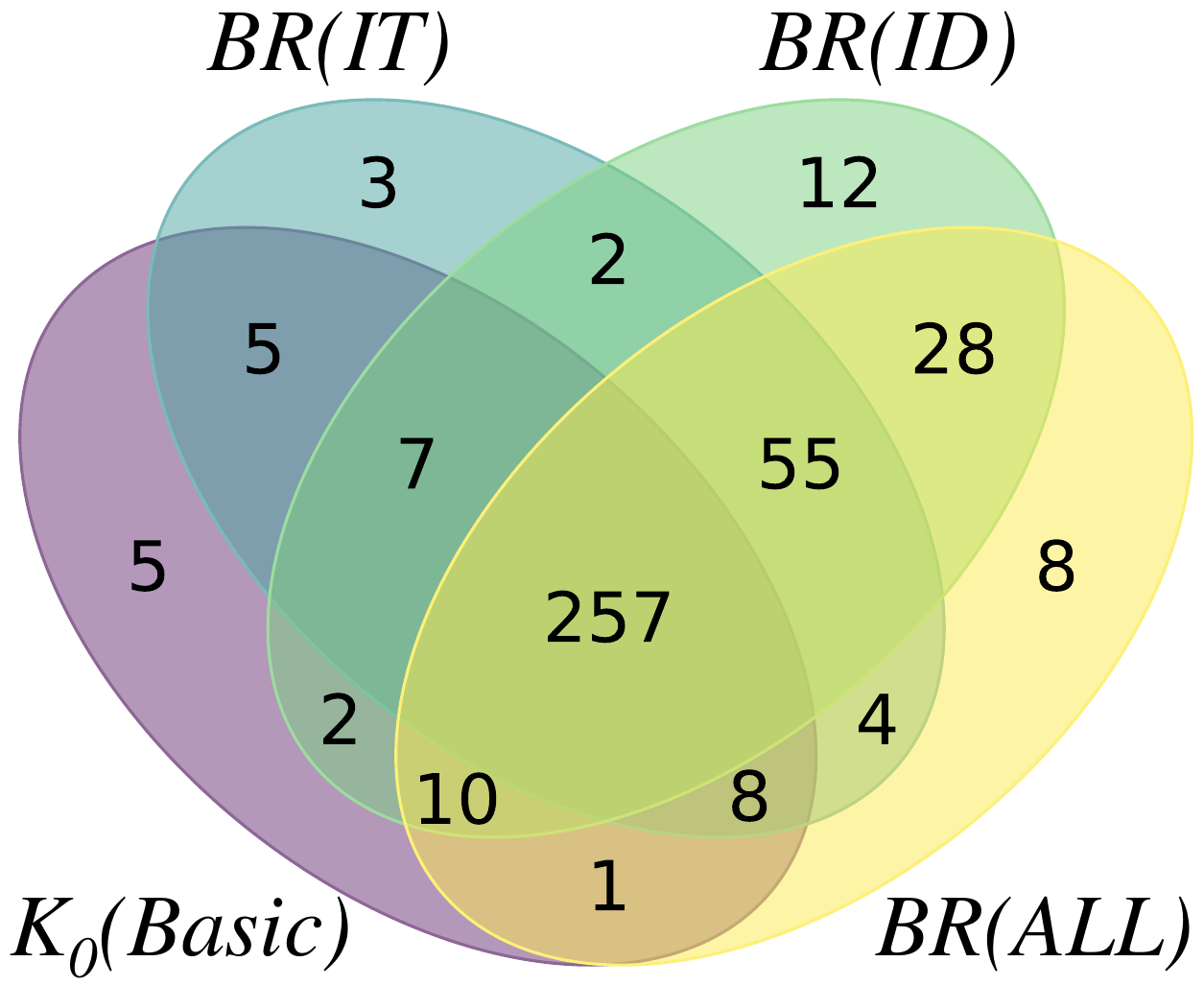}
        \caption{Venn diagram}
        \label{fig:venn-bp}
    \end{subfigure}
    \begin{subfigure}[t]{2.05in}
        \centering
        \includegraphics[width=\textwidth]{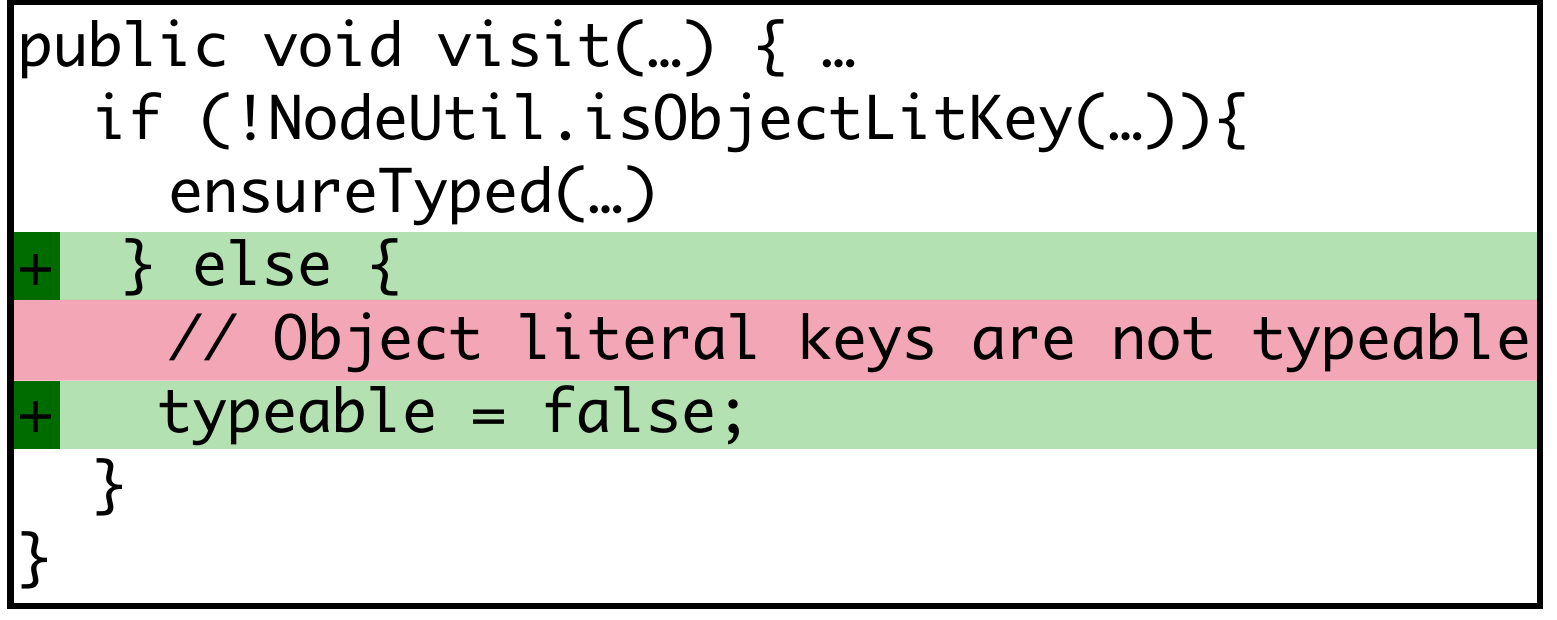}
        \caption{Closure-66 bug}
        \label{fig:closure-66-bug}
    \end{subfigure}
    \caption{The Venn diagram of plausible fixes over different setups and the bug Closure-66 which can only be fixed in \basic{}}
    \label{fig:EC-Ratio-upon-AFL}
\end{figure}

Noticing that since applying zero-shot learning achieves the optimal \aprperf{} among all the k-shot learning techniques as mentioned, we also adopt zero-shot learning in our \repairinfo{} evaluations and \basic{} as a baseline for a fair performance comparison. Table~\ref{tab:rq2-bpat-result} presents the APR results under different \repairinfo{} setups. We observe that using bug report-relevant setups significantly enhances the \aprperf{} of all models, i.e., the number of average \plsfix{} increases from 180 in \basic{} to 238 in \issuetitle{}, 270 in \issuedes{}, and 273 in \bugreport{}. %It should be noted that \issuetitle{}, on average, contains only 12 tokens, yet it resulted in a 32.2\% improvement in repair performance. Furthermore, we can find that all models demonstrated notable enhancements in the three distinct bug report-based settings, with improvements ranging from a 21.9\% increase for \codellama{} 7B using \issuetitle{}, to a remarkable 73.0\% for \codexedit{} in the \bugreport{} setting. 
On the other hand, while \bugreport{} achieves the optimal result, Figure~\ref{fig:venn-bp} also shows that it misses fixing 19 bugs which can be fixed in \basic{}. More specifically, we can observe that five bugs can only be fixed in \basic{} other than all the rest setups, such as Closure-66~\cite{Closure-66} in Figure~\ref{fig:closure-66-bug}. We find that to fix such a bug, a simple branch condition must be added to set \codeIn{typeable} false. In \basic{}, four models successfully fix this bug. However, in \issuetitle{}, \issuedes{}, and \bugreport{}, the focus is incorrectly placed on the issue described in the bug report, leading to inappropriate code logic changes. Consequently, none of the six models are able to fix the bug. %We also observe that inaccuracies in the bug report lead to a decline in the number of plausible patches for 43 bugs in the \bugreport{} setting compared to the \basic{} setting.

\mybox{Finding 4: While applying the bug report-relevant information significantly enhances the function-level \aprperf{}, it still misses fixing certain bugs which can be fixed by the baseline technique.}%  the variable quality of bug reports inadvertently reduces model repair efficiency for certain bugs, in some cases even leading to misdirected and unsuccessful repairs.}

We also attempt to investigate the performance impact from the \projinfo{} on the LLM-based function-level APR. Table~\ref{tab:rq2-bpat-result} shows that using \projinfo{} setups leads to an increase for all models, i.e., the average number of \plsfix{} rises from 180 in \kbasic{} to 185 in \buggycomment{}, 227 in \errormsg{}, 228 in \triggertest{}. Notably, \tec{} achieves an optimal average of 254 \plsfix{}, indicating the potential of leveraging as much \repairinfo{} as possible for enhancing the function-level \aprperf{}. Interestingly, unlike adopting the bug report-relevant information, all the bugs plausibly fixed in \basic{} can also be fixed by adopting the \projinfo{}. %More specifically, we can find that \gptturbo{} consistently obtains notable performance improvements under various \repairinfo{} settings, with the most substantial enhancement observed in the \tec{} setting, achieving up to 273 \plsfix{}. Such result indicates that by leveraging their learning power and comprehensive analytical capabilities, these models can identify the root cause of defective code based on \projinfo{} — including trigger tests, corresponding error messages, and natural language comments — thereby facilitating function-level repair.

% Table~\ref{tab:rq2-bpat-result} shows that \triggertest{}, \errormsg{}, and \buggycomment{} lead to an increase in average \plsfix{} by 26.7\%, 26.1\%, and 2.8\% respectively across all models, \yyy{unify the comparison style} compared to \basic{}.

% Additionally, we observe that while the average \plsfix{} generally increases from 180 in the \basic{} setting to 185 in the \buggycomment{} setting, the impact of \projinfo{} on repair performance varies across models. For instance, the \codexedit{} model shows a 6.9\% increase in \plsfix{}, whereas \codellama{} 7B experiences a 2.6\% decrease.

%\yyy{still not quite rigid settings considering that actually you have a lot of combinations.}
 
\mybox{Finding 5: Directly adopting trigger tests, error messages, and comments from buggy projects can also effectively advance the function-level \aprperf{}.}

%auto-ignore

\begin{table}[]
    \caption {APR results with fault location information}
    \label{tab:rq3-fl-boostup}
    \centering
    \setlength\tabcolsep{9pt}
    % \begin{adjustbox}{width=0.98\textwidth}
    \resizebox{0.95\columnwidth}{!}{%
    \begin{threeparttable}
    \begin{tabular}{p{2cm}lrrr}
    \toprule
    % \hline
\textbf{Sources}&\textbf{Settings} & \bf{w/o \textsuperscript{†}FL}& \bf{w/ \textsuperscript{†}FL} & \bf{Improvement}\\
    \midrule      
  %           Repair Scenarios                                  w/ FL   W/O FL  Boost Up
\bf{N.A.} &   \bf{\basic{}}                                 &   180  &   217  &   \bf{20.6\%}  \\
\cmidrule(lr){1-5}   
\multirow{3}{2cm}{\textbf{Bug Report Information}}
          &   \bf{\issuetitle{}}                            &   238  &   262  &   \bf{10.1\%} \\ 
          &   \bf{\issuedes{}}                              &   270  &   289  &   7.0\% \\ 
          &   \bf{\bugreport{}}                             &   273  &   291  &   6.6\% \\ 
    \cmidrule(lr){1-5}
    \multirow{4}{2cm}{\textbf{Project-specific Information}}
          &   \bf{\buggycomment{}}                          &   185  &   217  &   \bf{17.3\%} \\ 
          &   \bf{\errormsg{}}                              &   227  &   257  &   13.2\% \\ 
          &   \bf{\triggertest{}}                           &   228  &   246  &   7.9\% \\ 
          &   \bf{\tec{}}                                   &   254  &   272  &   7.1\% \\ 
        % \hline
        \bottomrule
        \end{tabular}
        \begin{tablenotes}
        \item[{}] \textbf{\textsuperscript{†}FL refers to fault location information.}
        % \item[{}] \textbf{\tec{}\textsuperscript{\textdaggerdbl} represents the collective usage of \triggertest{}, \errormsg{}, and \buggycomment{}.}
        \end{tablenotes}
        \end{threeparttable}
        % \end{adjustbox}
        }
\end{table}

    %       &   \bf{\issuetitle{}}                            &   241  &   266  &   \bf{10.3\%} \\ 
    %       &   \bf{\issuedes{}}                              &   274  &   292  &   6.3\% \\ 
    %       &   \bf{\bugreport{}}                             &   276  &   292  &   5.9\% \\ 
    % \cmidrule(lr){1-5}
    % \multirow{4}{2cm}{\textbf{Project-specific Information}}
    %       &   \bf{\buggycomment{}}                          &   189  &   221  &   \bf{17.8\%} \\ 
    %       &   \bf{\errormsg{}}                              &   229  &   260  &   13.2\% \\ 
    %       &   \bf{\triggertest{}}                           &   234  &   257  &   9.8\% \\ 
    %       &   \bf{\tec{}}                                   &   257  &   275  &   6.7\% \\ 

We further evaluate the performance impact and necessity of the statement-level fault location information in the function-level APR. We utilize the ground-truth statement-level fault location information following previous work~\cite{codexapr-xy} by labeling the corresponding buggy line with \codeIn{/*bug is here*/}. Specifically, the ground-truth fault locations are provided by the official \defectsj{} GitHub Repository~\cite{d4j-github}. To investigate the impact of the statement-level fault location information on the function-level APR, we calculate the average number of \plsfix{} generated by various models across different \repairinfo{} setups. 

% For clarity, we use the abbreviation ``FL'' to refer to statement-level fault localization information. Additionally, we apply ``w/'' and ``w/o'' to denote with and without fault localization information, respectively.

\begin{figure}[!htb]
    \centering
    \includegraphics[width=0.98\columnwidth]{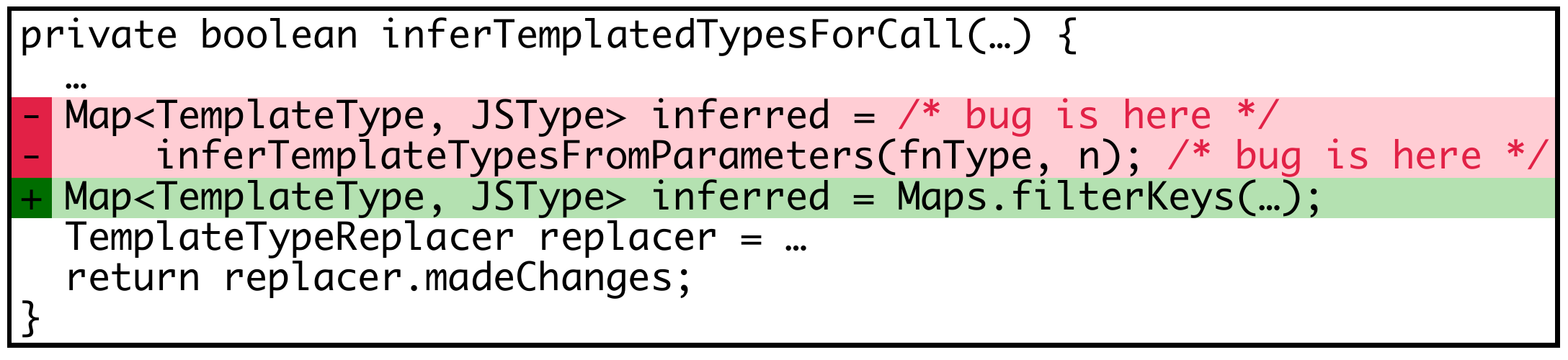}
    \caption{Statement-level fault location information misleads LLM, preventing the repair of the Closure-112 bug}
    \label{fig:fl-failure-case}
\end{figure}

From Table~\ref{tab:rq3-fl-boostup}, we can observe that while applying the statement-level fault location information enhances the repair performance, the extent of this improvement can be potentially compromised with the token number increase of the \repairinfo{}. For instance, while \basic{}$_{FL}$ achieves a performance improvement of 20.6\% compared to \basic{}, such an improvement shrinks to 6.6\% comparing \bugreport{}$_{FL}$ to \bugreport{} with averagely 246 tokens and 7.1\% comparing \tec{}$_{FL}$ to \tec{} with averagely 396 tokens. Moreover, we find that 14 bugs that are originally fixable without fault location information cannot be plausibly fixed across all setups and models when using fault location information. For instance in the Closure-112 bug~\cite{Closure-112} shown in Figure~\ref{fig:closure-66-bug} which demands multiple edits, a correct fix is achieved if the model reads the entire method, thus comprehending the necessity of adding \codeIn{Maps.filterKeys} to check if each key (of the \codeIn{TemplateType} type) exists in the key collection. However, with the fault location information, the attention of the model becomes disturbed, consequently over-focusing on the \codeIn{Map<TemplateType, JSType> inferred} code block and making extensive but ineffective modifications.

\mybox{Finding 6: The statement-level fault location information effectively enhances the \aprperf{}. As the token number of \repairinfo{} increases, the extent of the improvement can be potentially compromised.}%\jh{but not for \projinfo{}, the risk of overgeneralization exists.}}

\section{discussion}

\begin{figure*}[!htb]
    \centering
    \includegraphics[width=0.95\textwidth{}]{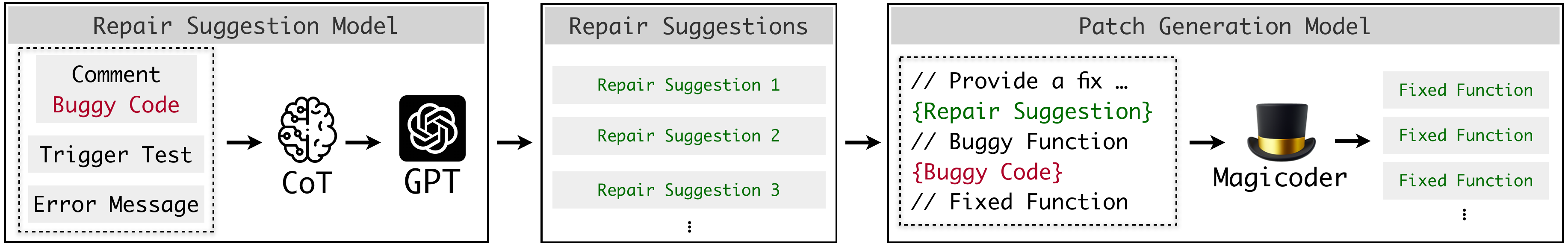}
    \caption{The \srepair{} framework}
    \label{fig:approach-framework}
\end{figure*}

\subsection{Bug report}

While the bug reports associated with carefully evaluated projects like \defectsj{} are generally of high quality where their effectiveness can be shown in our evaluation results, they nonetheless include instances of inaccuracies~\cite{ifixr}. %, e.g., the number of plausible patches for 43 bugs in \bugreport{} is fewer than in \basic{}. This issue is magnified 
Specifically, real-world bug reporting is filled with a significant volume of reports that are invalid, irreproducible, incomplete, or outright misleading~\cite{bpQuiality1, bpQuiality2, r2fix}. Moreover, the process of generating bug reports is manual and labor-intensive, in contrast to the APR techniques seeking to rectify software bugs autonomously, eliminating the need for human intervention~\cite{aprDefine}. Consequently, relying on bug reports for providing \repairinfo{} to advance the function-level APR may be inappropriate and impractical, especially when dealing with unknown faults. On the contrary, trigger tests~\cite{acs-xy,selfapr,spr} precisely identify the root cause of faults. Error messages~\cite{deepdelta,selfapr} can be automatically obtained from test outputs and  reveal the fault-triggering boundary conditions. Comments provide function descriptions added by developers~\cite{codexapr-xy}. These sources of information are more precise and cost-efficient compared to bug reports. Therefore, we recommend the utilization of \projinfo{} in LLM-based APR techniques to further improve repair performance.

% Due to the significant manpower required for developers to manually write bug reports and the varying quality of real-world bug reports, they are not suitable for automatically and efficiently providing reliable repair scenario information to models.

\subsection{Models for APR}
Although the \codellama{} models have gained a number of \plsfix{} in our study, we do observe abnormal behaviors of \codellama{}-based models in the patch generation of the function-level APR. When applied to the Java-based \defectsj{} dataset, \codellama{} models frequently generate patches with `\codeIn{[PYTHON]}' tags and Python code, e.g., producing 188,113 such patches in the \codellama{} 34B model. This issue was not prevalent in other models. Hence, we advocate using the high-performing \gptturbo{} and the open-source \magicoder{} models, both of which have shown superior capabilities in the APR task.

\section{Approach}

By far, we have demonstrated the power of adopting the \repairinfo{} in the function-level LLM-based APR, i.e.,  including such information in the repair prompt along with the buggy function under zero-shot learning. In this section, to further leverage the potential of the \repairinfo{}, we construct a novel function-level APR technique \srepair{} (referring to \textbf{S}uggestion \textbf{Repair}), which adopts a dual-LLM framework for advancing the \aprperf{}.

% which adopts a \dualllm{} framework and utilizes the learning power via Chain of Thought (CoT) technique~\cite{fewshotcot}, incorporating \projinfo{} (i.e., trigger tests, error messages, and comments), without the need for statement-level fault locations, in order to make APR technology more practical in the real-world application. %To this end, we first introduce the \srepair{} framework, then conduct an evaluation in function-level APR using both \defectsj{}~\cite{defects4j} and \quix{}~\cite{quix} datasets. Finally, we conduct a configuration study for the settings of \srepair{}, focusing on its applicability to different models and the costs associated with various settings.

\subsection{\srepair{} Framework}

% \note{change the fig}

\begin{figure}[!htb]
    \centering
    \includegraphics[width=0.9\columnwidth{}]{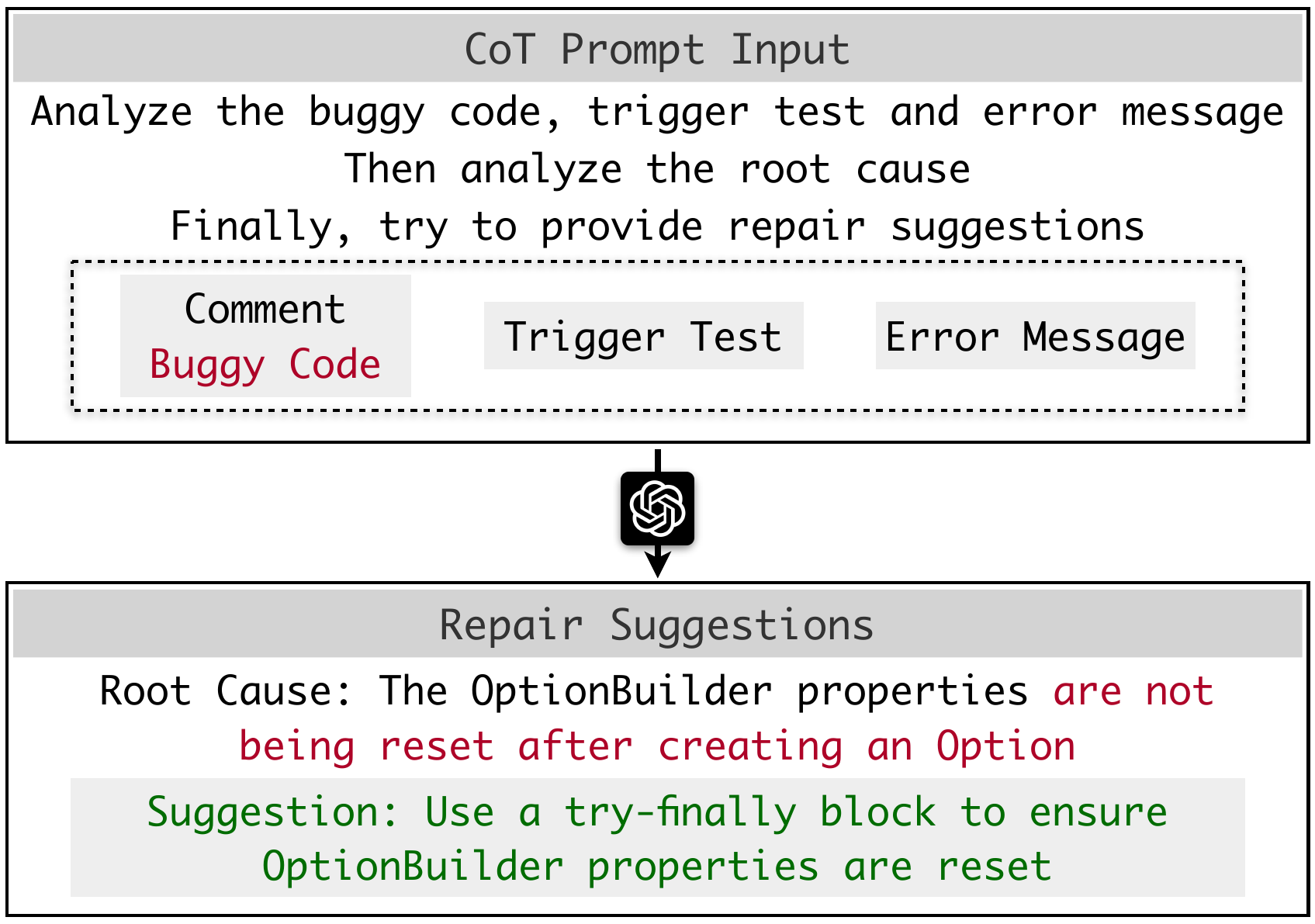}
    \caption{Chain of Thought example of Cli-26 Bug}
    \label{fig:cot-cli26}
\end{figure}

% For better leveraging the repair capability of the \repairinfo{}, we apply two LLMs for separating tasks as shown in Figure~\ref{fig:approach-framework}: the first analysis model utilizes the learning power of LLM by comprehensive analysis the \repairinfo{} via Chain of Thought technique~\cite{fewshotcot} and then gives the repair suggestions in nature language; the second code model exhibits its code generation capabilities to generate the entire patched function following the repair suggestion given by analysis model. 
Our \dualllm{} framework is shown in Figure~\ref{fig:approach-framework} where \srepair{} first adopts a \suggestionmodel{} which utilizes the learning power of LLM by comprehensively analyzing the \repairinfo{} via the Chain of Thought (CoT) technique~\cite{fewshotcot}. Then it provides repair suggestions in natural language. Next, \srepair{} adopts a \patchmodel{} which exhibits its code generation capabilities by generating the entire patched function following the repair suggestions. More specifically, we enable the CoT technique by prompting the LLM to first analyze the buggy function and \projinfo{}, then identify the root cause of the bug, and finally generate repair suggestions in natural language. For instance, as shown in Figure~\ref{fig:cot-cli26}, the \suggestionmodel{} first identifies the root cause of the Cli-26 bug~\cite{Cli-26}: `\textit{are not being reset after creating an Option}', and then generates the correct repair suggestion, `\textit{use a try-finally block}'. Finally, such a suggestion is fed to the \patchmodel{} for generating the patched functions.

\subsection{Evaluation}

%auto-ignore
\begin{table}[]
\caption{Statistics of \srepair{} Dataset}
\label{approach_dataset}
\centering
\setlength\tabcolsep{12pt}
\resizebox{0.95\columnwidth}{!}{%
\begin{tabular}{llr|rr}
\toprule
\textbf{Dataset}           & \textbf{Project} & \textbf{\# Bugs} & \textbf{\begin{tabular}[c]{@{}r@{}}SF\\ Bugs\end{tabular}} & \textbf{\begin{tabular}[c]{@{}r@{}}MF\\ Bugs\end{tabular}} \\ \midrule
\multirow{6}{*}{\textit{\textbf{Defects4j 1.2}}}  & Chart            & 25               & 16        & 9                                               \\
                                                  & Closure          & 140              & 105                          & 35     \\
                                                  & Lang             & 56               & 42                   & 14    \\
                                                  & Math             & 102               & 74                     & 28          \\
                                                  & Mockito          & 30               & 24               & 6       \\
                                                  & Time             & 22               & 16                 & 6     \\ \midrule
\multirow{11}{*}{\textit{\textbf{Defects4j 2.0}}} & Cli              & 30               & 28                          & 2       \\
                                                  & Codec            & 13               & 11                              & 2        \\
                                                  & Collections      & 2                & 1                       & 1    \\
                                                  & Compress         & 40               & 36                           & 4          \\
                                                  & Csv              & 13               & 12                 & 1       \\
                                                  & Gson             & 12                & 9            & 3      \\
                                                  & JacksonCore      & 18               & 13         & 5          \\
                                                  & JacksonDatabind  & 85               & 67                      & 18      \\
                                                  & JacksonXml       & 5                & 5            & 0      \\
                                                  & Jsoup            & 58               & 53             & 5     \\
                                                  & JxPath           & 14               & 10             & 4   \\ \midrule
\textbf{Overall}                                  & \textbf{}        & 665              & 522                           & 143                                                                                 \\ \bottomrule
\end{tabular}%
}
\end{table}

\subsubsection{Dataset}
We use the widely studied repair benchmark of \defectsj{}~\cite{defects4j} and \quix{}~\cite{quix}. Specifically, to extensively leverage \srepair{}'s ability in the function-level APR, we include all function-level bugs from \defectsj{} 1.2 and 2.0, thereby forming a dataset that comprises 522 single-function (SF) bugs and an additional 143 multi-function (MF) bugs, i.e., the bugs existing in multiple functions and requiring simultaneous edits on them for a fix, as shown in Table~\ref{approach_dataset}. Additionally, we also evaluate on the \quix{} dataset which is made up of 40 function-level buggy and fixed versions of classic programming problems in both Python and Java.

%auto-ignore
\begin{table*}[ht]
    \caption {Single-function APR result of \srepair{}}
    \label{tab:approach-sf-result}
    \centering
    \begin{adjustbox}{width=\textwidth}
    % \begin{threeparttable}
    \begin{tabular}{clrr|rrrr|rrrrr}
 
    \toprule
    % \hline
    \multirow{3}{*}{\textbf{Datasets}}& \multirow{3}{*}{\textbf{Project}}&\multicolumn{6}{|c|}{\textbf{Plausible Fixes}} & \multicolumn{5}{c}{\textbf{Correct Fixes}} \\
    \cmidrule{3-13}
&& \multicolumn{2}{|c|}{\bf{\tec{}}}  &\multicolumn{4}{c|}{\bf{\srepair{} Variant}}  &\multirow{2}{*}{\textbf{AlphaRepair}} & \multirow{2}{*}{\textbf{Repilot}} &\multirow{2}{*}{\textbf{FitRepair}} & \multirow{2}{*}{\textbf{\chatrepair}} & \multirow{2}{*}{\textbf{\srepaircmp}}\\
    \cmidrule(lr){3-4}  \cmidrule(lr){5-8} 
&  &\multicolumn{1}{|c}{\textbf{\gptturbo{}}} &\textbf{\magicoder{}} &\textbf{\srepair{}$_{2M}$} & \textbf{\srepair{}$_{2M+FL}$}  & \textbf{\srepairbsc{}}&\textbf{\srepaircmp}& \\
    % \hline
    \midrule
    \multirow{6}{*}{\textit{\textbf{\defectsj 1.2}}}
   %  Project       GPT3.5     MAGICODER     DUAL-LLM    DUAL-LLM FL  DUAL-LLM COT    PLAUSIBLE   CORRECT
  & Chart       &   \multicolumn{1}{|r}{12}     &   11  &   14  &   14  &   14  &   \bf{14}  &9&6&8&\textbf{15} &   13 \\   
  & Closure     &   \multicolumn{1}{|r}{40}     &   30  &   39  &   49  &   48  &   \bf{56}  &23&22&29&37 &   \textbf{47} \\ 
  & Lang        &   \multicolumn{1}{|r}{19}     &   25  &   27  &   29  &   29  &   \bf{32}   &13&15&19&21&   \textbf{26} \\ 
  & Math        &   \multicolumn{1}{|r}{48}     &   43  &   50  &   48  &   47  &   \bf{55}   &21&21&24&32&  \textbf{ 42} \\    
  & Mockito     &   \multicolumn{1}{|r}{8}      &   8   &   12  &   9   &   12  &   \bf{12}   &5&0&6& 6&   \textbf{11}\\ 
  & Time        &   \multicolumn{1}{|r}{7}      &   5   &   5   &   6   &   6   &   \bf{7}    &3&2&3&3&   \textbf{7} \\ 
  % \cdashline{1-13}
% \addlinespace
  % \multicolumn{2}{c}{\textbf{D4J 1.2 Total}}
  % &\textbf{D4J 1.2}& \multicolumn{1}{|r}{1}    & 1& 1&1& 1&1 & 1\\
 \midrule
    \multirow{11}{*}{\textit{\textbf{\defectsj 2.0}}}
  & Cli         &   \multicolumn{1}{|r}{16}     &   13  &   16  &   17  &   17  &   \bf{19}   &5&6&6&5&   \textbf{18} \\ 
  & Codec       &   \multicolumn{1}{|r}{8}      &   5   &   8   &   8   &   8   &   \bf{11}    &6&6&5& 8&   \textbf{11}\\ 
  & Collections &  \multicolumn{1}{|r}{0}       &   1   &   0   &   1   &   1   &   \bf{1}     &0&1&1&0&   \textbf{1} \\
  & Compress & \multicolumn{1}{|r}{21}          &   22  &   21  &   24  &   26  &   \bf{28}   &1&3&2&2&   \textbf{21} \\
  & Csv     & \multicolumn{1}{|r}{10}           &   9   &   10  &   9   &   10  &   \bf{11}    &1&3&2&3 &   \textbf{11}\\
  & Gson        & \multicolumn{1}{|r}{6}        &   8   &   7   &   7   &   7   &   \bf{9}      &2&1&1&3&   \textbf{8}\\
  & JacksonCore    & \multicolumn{1}{|r}{9}     &   6   &   9   &   7   &   9   &   \bf{10}   &3&3&3&3 &   \textbf{10} \\
  & JacksonDatabind    & \multicolumn{1}{|r}{30} &  28  &   39  &   38  &   39  &   \bf{45}    &8&8&10&9 &   \textbf{33}\\
  & JacksonXml      & \multicolumn{1}{|r}{3}    &   1   &   1   &   3   &   1   &   \bf{3}     &0&0&0&1 &   \textbf{2} \\
  & Jsoup       & \multicolumn{1}{|r}{34}       &   35  &   33  &   35  &   35  &   \bf{39}    &9&18&13&14 &   \textbf{35}\\
  & JxPath      & \multicolumn{1}{|r}{2}        &   4   &   4   &   5   &   4   &   \bf{5}     &1&1&1&0 &   \textbf{4} \\
 % \cmidrule(lr){1-9} 
 
 % \multicolumn{2}{c}{\textbf{D4J 2.0 Total}}
 % \cdashline{1-9}
 % \addlinespace
% &\textbf{D4J 2.0} &  \multicolumn{1}{|r}{1}   & 1& 1&1 &1 & 1&1 \\
   % \cmidrule(lr){1-9}
   \midrule
   % &\textbf{Overall}
  
   \multicolumn{2}{c}{\textbf{D4J 1.2 Total}} &\multicolumn{1}{|r}{134} &122 & 147& 155&156 & \bf{176}&74&66&89&114 &\textbf{146} \\
   % \midrule
   \multicolumn{2}{c}{\textbf{D4J 2.0 Total}} &\multicolumn{1}{|r}{139} &132 & 148& 154&157 & \bf{181}&36&50&44&48 &\textbf{154} \\
   \midrule
    \multicolumn{2}{c}{\textbf{Overall}} &\multicolumn{1}{|r}{273} &254 & 295& 309&313 & \bf{357}&110&116&133&162 &\textbf{300} \\
    \bottomrule
    
        \end{tabular}
        % \begin{tablenotes}
        % \item[{}] \textbf{Any content?}
        % \end{tablenotes}
        % \end{threeparttable}
        \end{adjustbox}
\end{table*}

\subsubsection{Implementation}

% In the implementation of \srepair{}, \gptturbo{} serves as the \suggestionmodel{}, utilizing Chain of Thought (CoT)~\cite{fewshotcot,zeroshotcot} prompting with trigger tests, error messages, and buggy functions with corresponding comments to analyze and then generate repair suggestions. For \patchmodel{}, \magicoder{} receives these repair suggestions for further generating patched functions. Specifically, we take \gptturbo{} as the \suggestionmodel{} because it not only demonstrates the strongest analytical and coding capabilities in our study, particularly in the \tec{} setting, but also possesses superior natural language generation abilities, i.e., it can generate repair suggestions in natural language, which is a capability the code model lacks. For the code model, we adopt \magicoder{} due to its cost-efficiency, as it has only a 7B parameter size, and its promising code generation ability, as demonstrated in our study, which is only slightly inferior to the best-performing model \gptturbo{}. 

In the \srepair{} implementation, \gptturbo{} acts as the \suggestionmodel{} due to its superior analytical, coding, and natural language generation abilities, especially in \tec{}. \magicoder{} is adopted as the \patchmodel{} due to its cost-effectiveness and competent code generation ability. Notably, for each repair suggestion, \srepair{} generates \suggestionpatch{} patched functions via the \patchmodel{}. We set the sample size of \srepair{} 200 (denoted as \srepairbsc{}) for comparing with previous APR results in our study section and \srepairsample{} (denoted as \srepaircmp{}) for fair comparisons with previous APR techniques~\cite{fitrepair, coconut-xy, alpharepair, Repilot}. %we specifically conduct a \srepair{} version with a sample size of \srepairsample{}, referred to as . 
Similar to prior work~\cite{alpharepair, cure, dlfix}, we additionally add an end-to-end time limit of 5 hours to fix one bug. Moreover, to better repair the Python bugs in \quix{}, we replace the Java comment symbol `\codeIn{//}' in the input APR prompt with the Python comment symbol `\codeIn{\#}'. It should be noted that \srepair{} does not require statement-level fault location information. For the rest setups, we follow our study section. Due to the page limit, we show the experimental results under different configurations and costs of \srepair{} in our GitHub page~\cite{githubrepo}.
% \note{CoT setting details}

\subsubsection{Evaluation Metrics}
Following our study section, we utilize plausible patches to reflect the repair performance. %Following standard practices in the APR research, we manually inspect each plausible patch for semantic equivalency\cite{llm4apr-xy,coconut-xy,alpharepair, angelix-xy, GandV} for determining the correct patches. 
Furthermore, following standard practices in the APR research, we manually inspect each plausible patch for semantic equivalency~\cite{llm4apr-xy,coconut-xy,alpharepair, angelix-xy, GandV} to determine the correct patches. Due to the intensive manual efforts involved in patch inspection, we conduct a cross-validation with three authors in order to filter out the correct patches generated by \srepaircmp{}.
% It should be noted that \srepair{}'s variants generate more than 10,000 unique plausible patches.

\subsubsection{Compared Techniques}
We adopt four recent SOTA LLM-based APR techniques: AlphaRepair~\cite{alpharepair}, Repilot~\cite{Repilot}, FitRepair~\cite{fitrepair}, and \chatrepair{}~\cite{chatrepair-xy}. We also adopt \gptturbo{}$_{\tec{}}$ and \magicoder{}$_{\tec{}}$ as baselines with the same \repairinfo{} and models used in \srepair{} for studying the effectiveness of our \dualllm{} CoT framework. We also form two \srepairbsc{} variants: \srepair{}$_{2M}$ with \dualllm{} only, i.e., directly generating repair suggestions without CoT, and \srepair{}$_{2M+FL}$ with additional statement-level fault location information for comparison.
% \dualllm{} with additional statement-level fault location information srepair{}$_{2LLM+FL}$ for comparison. 

% In particular, AlphaRepair is a state-of-the-art LLM-based repair technique by applying pre-trained CodeBERT model with cloze-style APR. Repilot is an innovative LLM-based APR technique designed to address the limitations of traditional LLMs in understanding programming semantics, thereby enhancing the synthesis of valid patches. FitRepair is a state-of-the-art LLM-based APR technique which combines LLMs with domain-specific fine-tuning and prompting strategies, fully automating the plastic surgery hypothesis. It should be noted that all these LLM-based APR techniques are designed for single-line or hunk-level bug fixing, while \srepair{} are designed for the practical function-level APR. 
% \note{wait the data}

%auto-ignore
\begin{table}[]
\caption{Correct fixes on \quix{} datasets}
\label{quix-result}
\centering
\resizebox{0.99\columnwidth}{!}{%
\begin{tabular}{lrrrrr}
\toprule
\textbf{QuixBugs} & \textbf{\srepaircmp{}} & \textbf{\srepairbsc{}} &\textbf{\chatrepair} & \textbf{AlphaRepair} \\

\midrule

\textbf{Python}     &\textbf{40}     &40         &40 &27 \\
\textbf{Java}       &\textbf{40}     &40         &40 &28 \\
\bottomrule
\end{tabular}%
}
\end{table}

\subsubsection{Result analysis}

% To evaluate \srepair{}, we include the \tec{} settings of \gptturbo{} and \magicoder{} as baselines, since \srepair{} utilizes the same models and \repairinfo{}, i.e., trigger tests, error messages and comments. Furthermore, to study \srepair{}'s effectiveness, we have designed different variants. `\dualllm{}' represents the collaborative work of two LLMs: the analysis model initially generates repair suggestions, which are then used by the code model to create the fixed function. Although \srepair{} does not adopt statement-level fault localization information, we introduced a variant named `\dualllm{} FL' to explore the impact of statement-level information on performance, compared to standard \srepair{}. Finally, `\dualllm{} CoT' denotes the ultimate \srepair{} setting that incorporates both the \dualllm{} approach and the CoT (Chain of Thought) prompting technique.

 % Specifically, in the `\srepair{} Variant' column, `\dualllm{}' denotes the collaborative work of two LLMs: the analysis model first generates repair suggestions, followed by the code model creating the fixed function based on these suggestions. Additionally, in the `\dualllm{} FL' variant, similar to the section \note{}, we integrate statement-level fault localization information to \dualllm{} by labeling the defect locations in the buggy function. The `\dualllm{} CoT' represents the \srepair{} setting that employs both the \dualllm{} approach and the CoT (Chain of Thought) prompting technique.

% and obtains 128 new correct fixes
Table~\ref{tab:approach-sf-result} presents the APR results for single-function bugs in the \defectsj{} dataset. Surprisingly, we find that \srepaircmp{} outperforms all previous LLM-based APR techniques by at least 85\%. Specifically, we can observe that 68.4\% of single-function bugs (357) in \defectsj{} can be plausibly fixed, and even 57.5\% of bugs (\totalcorrect{}) can be correctly fixed by \srepaircmp{}. Such surprising results indicate that \srepair{} is capable of fixing a significant number of real-world complicated bugs in the function-level APR. Notably, repairing 300 single-function bugs with \srepair{} costs only \$8.6, averaging \$0.029 per correct fix, demonstrating its efficiency as an LLM-based APR technique.

\begin{figure}
    \centering
    \begin{subfigure}[t]{1.6in}
        \centering
        \includegraphics[width=\textwidth]{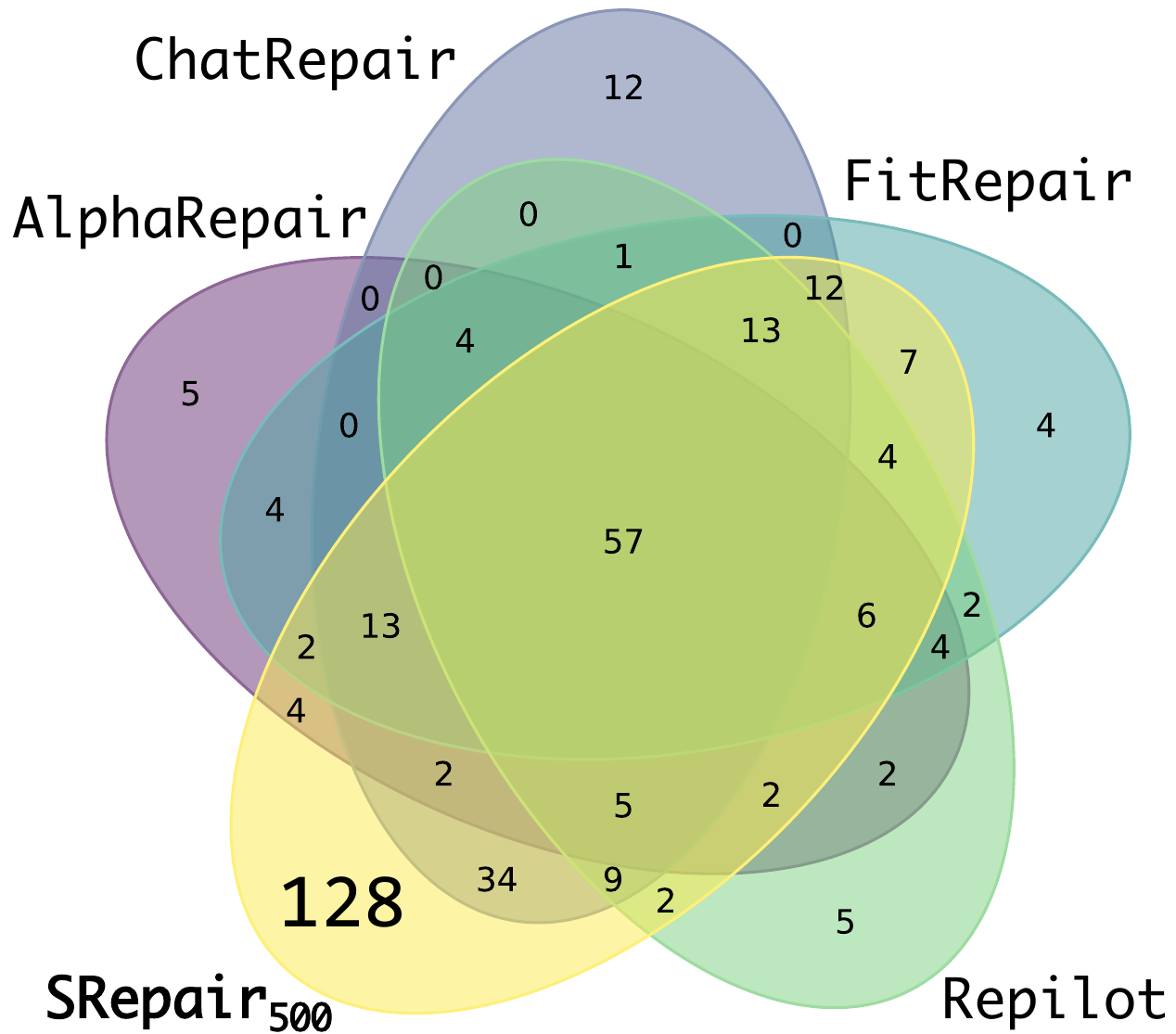}
        \caption{Single-function dataset}
        \label{fig:venn-srepair-all}
    \end{subfigure}
    \begin{subfigure}[t]{1.6in}
        \centering
        \includegraphics[width=\textwidth]{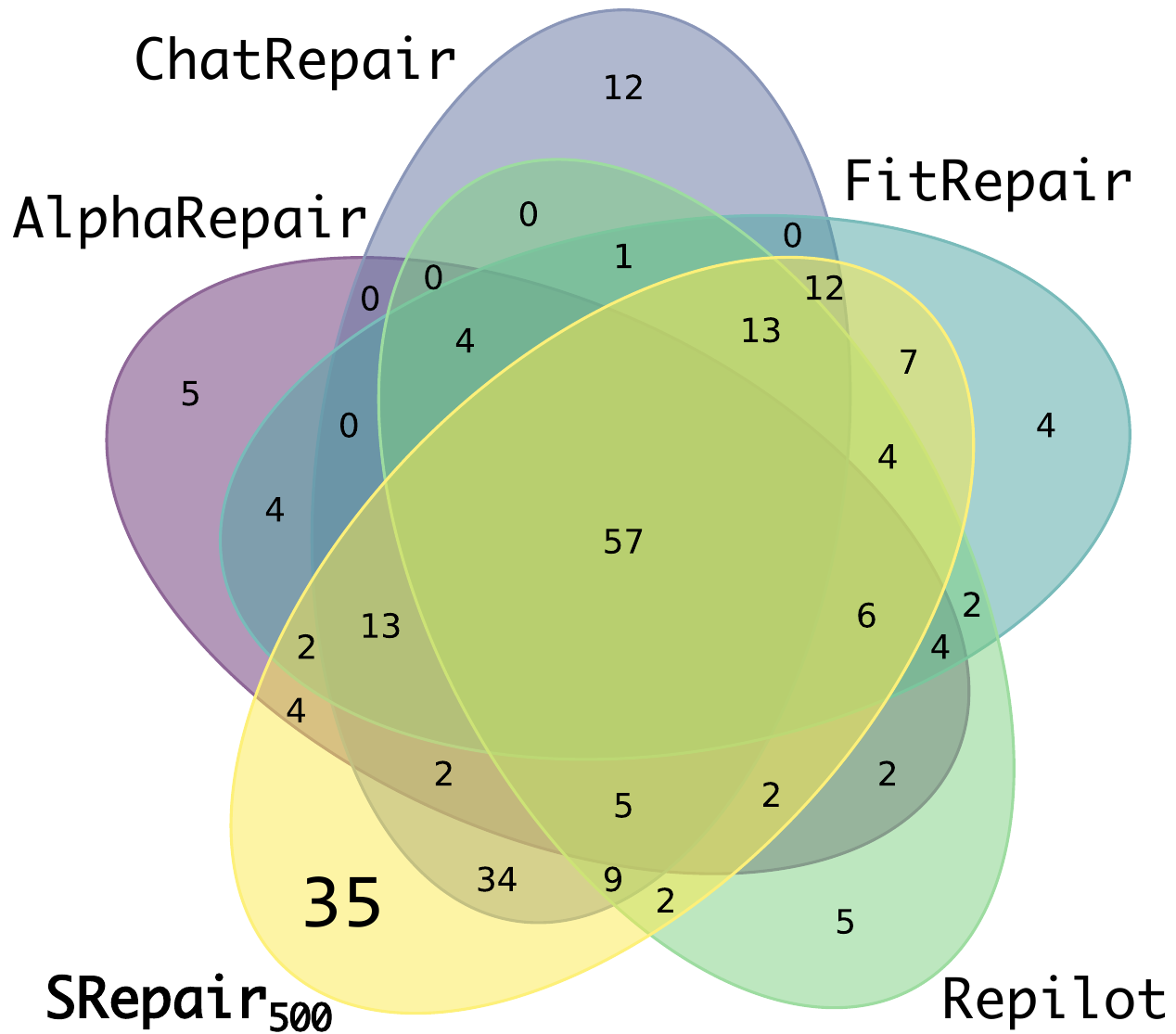}
        \caption{Studied baselines dataset}
        \label{fig:venn-srepair-limited}
    \end{subfigure}
    \caption{Bug fixes Venn diagram of \srepaircmp{} with studied baselines}
    \label{fig:venn-srepair}
\end{figure}

Moreover, as in Figure~\ref{fig:venn-srepair-all}, \srepaircmp{} correctly fixes 128 out of 522 single-function bugs which cannot be fixed by any of the baseline LLM-based APR techniques adopted in this paper. Interestingly, Figure~\ref{fig:venn-srepair-limited} shows that \srepaircmp{} also significantly outperforms the state-of-the-art APR baselines, correctly fixing 35 unique bugs that all other baselines failed to fix in their studied bugs. Such a result indicates that \srepair{} not only expands the repair task scope to the more practical function-level APR but also achieves remarkable \aprperf{} without the need for statement-level fault location information. Table~\ref{quix-result} shows that \srepaircmp{} successfully fixes all bugs in the \quix{} dataset, indicating its superior capability for diverse programming languages.

% where we merge their single-line or hunk-level repair results as the function-level results. 
% Moreover, as illustrated in the Venn diagram in Figure~\ref{fig:venn-srepair-limited}, \srepaircmp{} significantly outperforms the state-of-the-art APR baselines on the same dataset, i.e., the bugs were previously attempted to be repaired by the studied baselines, correctly fixing 35 unique bugs that other baselines failed to address. For all the single-function bugs in \defectsj{}, \srepaircmp{} can achieve 128 unique correct fixes as shown in Figure~\ref{fig:venn-srepair-all}, surpassing the four baselines.
% It should be noted that while all studied baselines utilize statement-level defect locations and conduct repairs at the line/hunk level, \srepaircmp{} operates at the function level and achieves remarkable \aprperf{}. 
\begin{figure}[!htb]
    \centering
    \includegraphics[width=0.96\columnwidth{}]{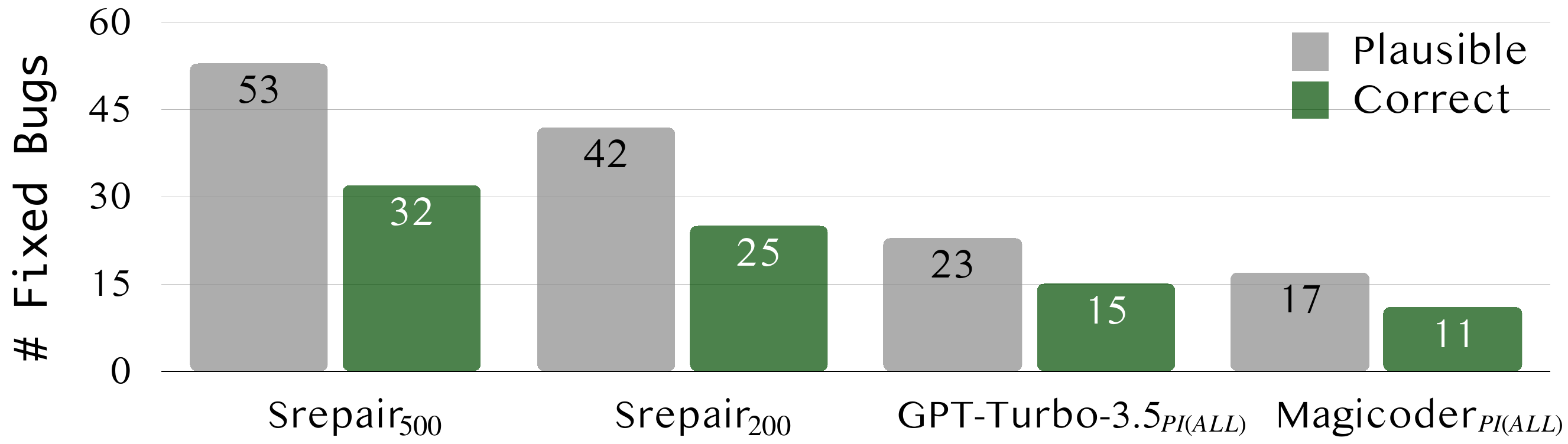}
    \caption{The APR results of the multi-function bugs in the \defectsj{} dataset}
    \label{fig:mf-result}
\end{figure}

\begin{figure}
    \centering
    \includegraphics[width=0.99\columnwidth]{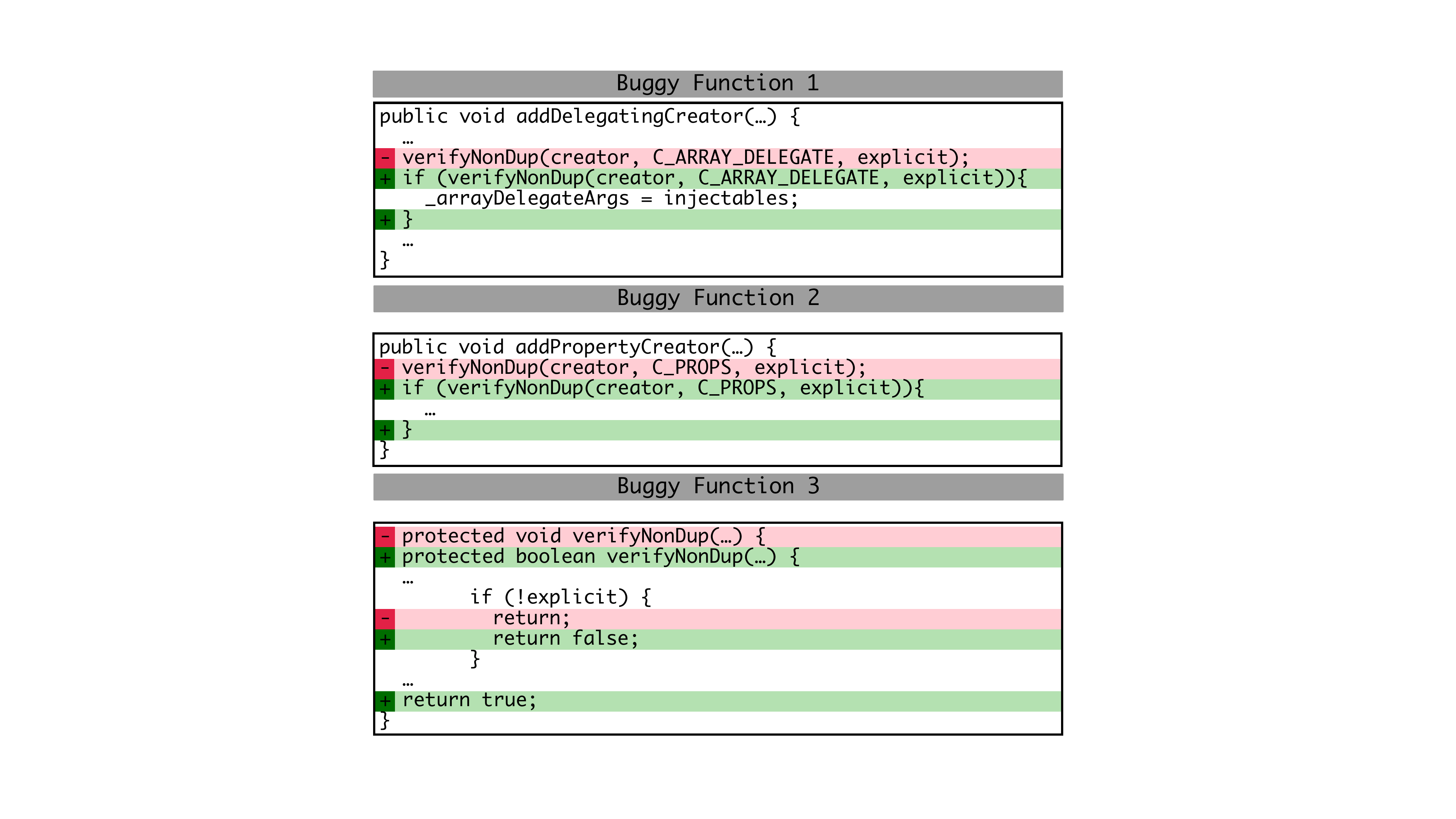}
    \caption{The multi-function bug JacksonDatabind-69~\cite{JacksonDatabind-69}}
    \label{fig:jd-69}
\end{figure}

We also evaluate how \srepair{} repairs complicated multi-function bugs shown in Figure~\ref{fig:mf-result} where we find that \srepaircmp{} (53 plausible fixes and \multifunctcorrect{} correct fixes) and \srepairbsc{} (42 plausible fixes and 25 correct fixes) both largely outperform \gptturbo{}$_{\tec{}}$ and  \magicoder{}$_{\tec{}}$. %While \srepair{} outperforms \gptturbo{}$_{\tec{}}$ by 14.7\% in terms of plausible single-function repairs, it significantly surpasses \gptturbo{}$_{\tec{}}$ by 82.6\% in generating plausible multi-function fixes and by 66.7\% in producing correct fixes. 
Interestingly, as shown in Figure~\ref{fig:jd-69} where Functions 1 and 2 require information from successfully running Function 3 to determine if they should execute subsequent statements. This poses a significant challenge for APR techniques, as they need to simultaneously alter the return type of Function 3 to boolean and adapt the function calls in Functions 1 and 2. \srepair{} successfully identifies such a complex function call and generates the correct fix, indicating the power of \srepair{} on complicated multi-function faults, which, to our best knowledge, is the first time achieved by any APR technique ever.

%While \srepaircmp{} yields the most promising results, \srepairbsc{} still achieves 87.7\% (313 out of 357) of plausible fixes compared to \srepaircmp{}. Additionally, 
We further find that \srepair{}$_{2M}$ outperforms both \gptturbo{}$_{\tec{}}$ and \magicoder{}$_{\tec{}}$ by 8.1\% and 16.1\% in terms of the number of plausible fixes respectively. Furthermore, leveraging CoT technique achieves even better result (313 plausible fixes) than incorporating statement-level fault localization information (309 plausible fixes). %improves \dualllm{}'s repair performance by 4.7\%, while leveraging CoT technique achieves even better results, with 309 plausible fixes in \srepair{}$_{2M+FL}$ compared to 313 in \srepairbsc{}. 
Such results indicate the effectiveness of our \dualllm{} framework and CoT mechanism in \srepair{}.

\section{Threats to validity}

\noindent \textbf{Threats to internal validity.}
One potential threat arises from our manual validation process, which differentiates between plausible patches and those that are semantically correct. To address this concern, three authors cross-validated the plausible patches of \srepaircmp{} by comparing them to those created by developers (the plausible patches generated by other techniques are mostly subsets). %Furthermore, the correct patches and our experiment code are publicly available for review and evaluation~\cite{githubrepo}.

Another threat is the potential for data leakage if the developer patches were included in the original training data. To address this, we examined all the patches generated in our study and by \srepaircmp{} in the \defectsj{} dataset. Among the total plausible patches produced in our study, only 7.4\textperthousand{} are identical to the developer patches. Similarly, for the plausible patches generated by \srepaircmp{}, only 1.5\textperthousand{} match the developer patches. Such overlapped patches pose almost no impact on our experimental results. %Furthermore, \srepaircmp{} successfully fixed 129 new single-function bugs in both \defectsj{} 1.2 and 2.0 datasets compared to four state-of-the-art LLM-based APR techniques. It should be noted that, given we utilize six different LLMs, completely solving the data leakage issue by re-training all these models is infeasible.
% 30.1% 15.5% 5‱ 0.8‱

An additional threat lies in the trigger tests adopted in \srepair{} where the LLMs might have recognized the trigger tests and manipulated them to pass all tests, creating seemingly plausible patches. Our \srepair{}'s \dualllm{} mechanism effectively mitigates this threat, as the \suggestionmodel{} only suggests bug fixes without trigger test information, keeping the \patchmodel{} isolated from such data. %Additionally, three developers manually reviewed all plausible patches generated by \srepaircmp{} and confirmed that none of the patches passed tests by merely handling trigger test cases.

\noindent \textbf{Threats to external validity.}
The main threat to external validity lies in our evaluation datasets used which may not well generalize our experimental results. To mitigate this, we evaluate our approach on both the popular \defectsj{} 1.2 and 2.0 datasets where we include all their single-function bugs in our study. Furthermore, we extend our investigation to multi-function bugs in our \srepair{} evaluation. We also evaluate \srepair{} on the \quix{} datasets, which contain both Java and Python bugs, to validate its generalizability.

\noindent \textbf{Threats to construct validity.} The threat to construct validity mainly lies in the metrics used. To mitigate this, we adopt the widely-used plausible patches along with their distributions. We also use correct fix to evaluate our approach \srepair{}. %  To reduce this threat, we manually inspect all plausible patches generated by \srepaircmp{} to ensure the accuracy of the patch results. It should be noted that in the section of our study, over 10 million patches were generated and validated, resulting in more than 50,000 unique plausible patches. Due to the sheer volume, manual inspection of these plausible patches to obtain correct patches is not feasible.
%auto-ignore

\section{conclusion}

In this paper, we conduct the first comprehensive study on the function-level LLM-based APR. % including investigating the effect of the few-shot learning and the \repairinfo{}. 
Our study reveals that LLMs with zero-shot learning are powerful function-level APR techniques. Moreover, directly applying the \repairinfo{} to LLMs significantly increases the function-level \aprperf{}. Inspired by our findings, we design a \dualllm{} framework utilizing Chain of Thought technique, named \srepair{}, which achieves remarkable \aprperf{} by correctly fixing \totalcorrect{} single-function bugs in the \defectsj{} dataset, surpassing \chatrepair~\cite{chatrepair-xy} by 85\% and Repilot~\cite{Repilot} by 1.59$\times$. Notably, \srepair{} successfully fixes \multifunctcorrect{} multi-function bugs, which is the first time achieved by any APR technique ever to our best knowledge.

\section*{Data Availability}
The data and code are available at \git{}~\cite{githubrepo} for public evaluation.

% \newpage
% \clearpage
% \renewcommand*{\bibfont}{\scriptsize}
% \renewcommand*{\bibfont}{\footnotesize}
% \bibliographystyle{IEEEtran}
\bibliographystyle{ACM-Reference-Format}
\bibliography{llm4apr}
\end{document}